\documentclass[prb,twocolumn]{revtex4}
\usepackage{amssymb}
\usepackage{amsmath}
\usepackage{graphicx}

\begin{document}

\author{F. Mancini}
\title{New perspectives on the Ising model}
\address{Dipartimento di Fisica ''E.R. Caianiello" - Laboratorio
Regionale SuperMat, INFM \\ Universit\`{a} degli Studi di Salerno,
I-84081 Baronissi (SA), Italy}

\begin{abstract}
The Ising model, in presence of an external magnetic field, is
isomorphic to a model of localized interacting particles satisfying
the Fermi statistics. By using this isomorphism, we construct a
general solution of the Ising model which holds for any
dimensionality of the system. The Hamiltonian of the model is solved
in terms of a complete finite set of eigenoperators and eigenvalues.
The Green's function and the correlation functions of the fermionic
model are exactly known and are expressed in terms of a finite small
number of parameters that have to be self-consistently determined.
By using the equation of the motion method, we derive a set of
equations which connect different spin correlation functions. The
scheme that emerges is that it is possible to describe the Ising
model from a unified point of view where all the properties are
connected to a small number of local parameters, and where the
critical behavior is controlled by the energy scales fixed by the
eigenvalues of the Hamiltonian. By using algebra and symmetry
considerations, we calculate the self-consistent parameters for the
one-dimensional case. All the properties of the system are
calculated and obviously agree with the exact results reported in
the literature.
\end{abstract}

\date{\today}
\maketitle

\section{Introduction}

It is really very hard to say something new on the Ising model. The model,
originally proposed by Lenz\cite{Lenz20} in 1920, was exactly solved for the
case of an infinite chain by Ising\cite{Ising25} in 1925. Since then,
thousand and thousand of articles and several books have been published on
the subject. The reason is that the model is very simple, but still can be
considered as the prototype for systems subject to second order phase
transitions and can be effectively used for studying critical phenomena.
Moreover, the model and its generalizations and modifications can also be
used for studying a large variety of physical systems. We do not attempt to
summarize the enormous work done in these 80 years; it would go well beyond
the purpose of this article. An excellent historical presentation of the
Ising model can be found in Ref.~\onlinecite{Brush67}, although it is old
and obviously not updated. With no pretension of being exhaustive and
complete, we here summarize the principal approaches used in these 80 years.

A basic tool is the transfer matrix method\cite
{Kramers41,Kramers41a,Montroll41,Montroll42}. By means of this approach,
Onsager\cite{Onsager44} in 1944 succeed to give an exact solution of the
model for a cubic two-dimensional lattice in absence of external magnetic
field. The theory of spinors and Lie algebra was used to simplify the
Onsager solution \cite{Kaufman49,Newell53}. Among the exact results for the
two-dimensional case, the calculation of the magnetization\cite{Yang52} and
the writing of the spin correlation function in the form of a Toeplitz
determinant\cite{Montroll63} have to be mentioned. Other simplifications of
the Onsager solution have been obtained by means of the Jordan-Wigner
transformation and fermionization methods \cite{Schultz64,Lieb61,Lieb62}.
Different approaches are based on combinatorial methods\cite
{Kac52,Vdovichenko65a,Vdovichenko65b} and pfaffian methods\cite
{Hurst60,Green64,Kasteleyn63,McCoy73}. More recent approaches have seen the
Ising Hamiltonian expressed as a Gaussian Grasmannian action\cite
{Samuel80,Nojima98}. Along this line, use of operatorial symmetries that
simplify the algebra of the transfer matrix has led to the calculation of
the partition function for a large class of lattices\cite
{Plechko85,Plechko88}.

Many approximation methods have been used with the goal of obtaining an
expression for the partition function valid over a large temperature range:
mean field theory, Bethe approximation\cite{Bethe35}, cluster variational
methods\cite{Kikuchi51}, Monte Carlo simulations, series expansions. The
spin correlation functions have been studied at the critical temperature\cite
{Kadanoff69} and in the asymptotic region\cite{Wu66,Au-Yang77}. To study
critical phenomena and critical indices, tools like series expansions \cite
{Kirkwood38,Domb60,Domb65}, scaling\cite{Patashinskii66,Widom65,Kadanoff66},
renormalization group theory\cite{Wilson71,Wilson75} have been used.

In spite of the tremendous work done, many problems remain unsolved. The
exact partition function in a finite magnetic field is still unknown for
dimensions higher than one. Very few exact results have been obtained for
the three-dimensional model. There is no exact solution for the two-layer
Ising model either. Most of all, a general approach which works in all
dimensions and under general boundary conditions, although in some
approximation, is needed.

In a recent work\cite{Mancini05}, we have shown that there is a
large class of models which are exactly solvable in terms of a
finite number of parameters that have to be self-consistently
calculated. The purpose of the present paper is to apply the method
proposed in Ref.~\onlinecite{Mancini05} to the Ising model and to
show that an exact solution of the model does exist for any
dimension. In Section 2, we introduce the Ising model for a $d$
-dimensional cubic lattice and show that the model is isomorphic to
a system of localized spinless interacting particles, satisfying the
Fermi statistics. In Section 3, the Hamiltonian of the latter model
is solved, that is, a complete finite set of eigenoperators and the
relative eigenvalues are determined. Then, as shown in Section 4,
the exact form of the retarded Green's function (GF) and of the
correlation function (CF) can be obtained. In Section 5, we derive a
set of equations for determining the charge/spin correlation
functions. As the composite operators do not satisfy a canonical
algebra, the GF, the CF and the charge/spin correlation functions
depend on a set of internal parameters not calculable by the
dynamics. For the one-dimensional case, by means of the composite
operator method\cite {Mancini03,Mancini03a,Mancini04}, we calculate
these internal parameters (Section 6) and the charge/spin
correlation functions (Section 7). Although obvious, it is worth
noticing that all the results reproduce the exact solution known in
the literature.

What are the advantages of the present method and what is new in the
context of the Ising model? We present a new scheme of calculations
for treating the model. The scheme is general and can be applied to
any dimension. In the framework of this scheme we show that the
model is always solvable for all dimensions. The energy spectra of
the system are known. In the one dimensional case we show that the
energy scales determined by the spectra rule the behavior at the
critical temperature. It is reasonable to expect that this is true
also for higher dimensions. General relations among different spin
correlation functions have been obtained. These are exact relations
and might be used to check the consistency of some approximate
treatments or numerical calculations. In order to get quantitative
results for the cases of two and three dimensions we have to
determine a finite small number of parameters. All the properties of
the Ising model, the magnetization, the thermodynamical quantities,
the spin correlation functions, depend on these parameters that have
to be self-consistently determined. By using algebra and symmetry
considerations we calculate these parameters for the case $d=1$.
Extension of the calculations to higher dimensions is under
investigation.

\section{The Ising model}

The Ising model, in presence of an uniform external magnetic field $h$, is
described by the following Hamiltonian
\begin{equation}
H_{Ising}=\sum\limits_{\mathbf{ij}}{}J_{\mathbf{ij}}S(\mathbf{i})S(\mathbf{j}
)-h\sum\limits_{\mathbf{i}}{}S(\mathbf{i})  \label{2.1}
\end{equation}
$S(\mathbf{i})$ are spin variables, residing on a $d$-dimensional
Bravais lattice of $N$ sites spanned by the vectors
$\mathbf{R}_{i}=\mathbf{i}$ . The variables $S(\mathbf{i})$ takes
only two values: up or down, or more simply $S( \mathbf{i})=\pm 1$.
For a hypercubic lattice of lattice constant $ a$ with nearest
neighbor interactions, the exchange matrix $J_{\mathbf{ij}}$ is
given by
\begin{equation}
\begin{array}{l}
J_{\mathbf{ij}}=-2dJ\alpha _{\mathbf{ij}}\quad \quad \alpha
_{\mathbf{ij}}={ \frac{1}{N
}}\sum\limits_{\mathbf{k}}{}e^{\text{i}\mathbf{k}\cdot (\mathbf{R}
_{i}-\mathbf{R} _{j})}\alpha (\mathbf{k}) \\
\alpha (\mathbf{k})={\frac{1}{d}}\sum\limits_{n=1}^{d}{}\cos
(\mathbf{k} _{n}a)
\end{array}
\label{2.2}
\end{equation}
where $d$ is the dimensionality of the system and $\mathbf{k}$ runs over the
vectors in the first Brillouin zone. The exchange constant $J$ can be
positive or negative, and accordingly the coupling will be ferromagnetic or
antiferromagnetic. According to (\ref{2.2}), the Hamiltonian (\ref{2.1}) can
be rewritten as
\begin{equation}
H_{Ising}=-dJ\sum\limits_{\mathbf{i}}{}S(\mathbf{i})S^{\alpha }(\mathbf{i}
)-h\sum\limits_{\mathbf{i}}{}S(\mathbf{i})  \label{2.3}
\end{equation}
where
\begin{equation}
S^{\alpha }(\mathbf{i})=\sum\limits_{\mathbf{j}}{}\alpha
_{\mathbf{ij}}S( \mathbf{j})  \label{2.4}
\end{equation}
It is worth to recall that the Ising Hamiltonian (\ref{2.1}) is invariant
under the transformation
\begin{equation}
{S(\mathbf{i})\to -S(\mathbf{i})\qquad h\to -h}  \label{2.5}
\end{equation}
Let us consider a system of $N_{e}$ interacting spinless fermions
residing on the same lattice and let $c(i)$ and $c^{\dag }(i)$ be
the related annihilation and creation operators. These operators are
Heisenberg fields $ [i=(\mathbf{i},t)]$ satisfying canonical
anticommutation relations
\begin{equation}
\begin{array}{l}
{\{c(\mathbf{i},t),c^{\dag }(\mathbf{j},t)\}=\delta _{\mathbf{ij}}} \\
{\{c(\mathbf{i},t),c(\mathbf{j},t)\}=\{c^{\dag
}(\mathbf{i},t),c^{\dag }( \mathbf{j} ,t)\}=0}
\end{array}
\label{2.6}
\end{equation}
As a consequence of the algebra (\ref{2.6}), each site can be
occupied at most by a single particle. The occupation number of the
site $\mathbf{i}$, $ \nu (i)=c^{\dag }(i)c(i)$, takes only the
values 0 and 1. By taking into account two-body interactions, the
Hamiltonian for such a system reads as
\begin{equation}
H=-\sum\limits_{\mathbf{i}}{}\mu \nu
(i)+{\frac{1}{2}}\sum\limits_{\mathbf{ij }}{}V( \mathbf{i,j})\nu
(i)\nu (j)  \label{2.7}
\end{equation}
where $\mu $ is the chemical potential and $V(\mathbf{i,j})$ is the
potential. This model Hamiltonian can be connected to the Ising model by
defining
\begin{equation}
\nu (i)={\frac{1}{2}}[1+S(i)]  \label{2.8}
\end{equation}
It is clear that
\begin{equation}
\begin{array}{l}
\nu {(i)=0\quad \Leftrightarrow \quad S(i)=-1} \\
\nu {(i)=1\quad \Leftrightarrow \quad S(i)=+1}
\end{array}
\label{2.9}
\end{equation}
By substituting (\ref{2.8}) into (\ref{2.7}) and by considering only a
nearest-neighbor potential we can rewrite the Hamiltonian (\ref{2.7}) in the
following form
\begin{equation}
H=E_{0}-h\sum\limits_{\mathbf{i}}{}S(\mathbf{i})-dJ\sum\limits_{\mathbf{i}
}{}S(\mathbf{i} )S^{\alpha }(\mathbf{i})  \label{2.10}
\end{equation}
where we defined
\begin{equation}
\begin{array}{l}
{E_{0}=(-{\frac{1}{2}}\mu +{\frac{1}{4}}Vd)N} \\
{h={\frac{1}{2}}(\mu -Vd)} \\
{J=-{\frac{1}{4}}Vd}
\end{array}
\label{2.11}
\end{equation}
Hamiltonian (\ref{2.10}) is just the Ising Hamiltonian (\ref{2.3}) as we
have the equivalence
\begin{equation}
H_{Ising}=H-E_{0}  \label{2.12}
\end{equation}
The relation between the partition functions is
\begin{equation}
Z_{H}=e^{-\beta E_{0}}Z_{Ising}  \label{2.13}
\end{equation}
Then, the thermal average of any operator $A$ assumes the same value in both
models
\begin{equation}
\left\langle A(\nu )\right\rangle _{H}=\left\langle A(S)\right\rangle
_{Ising}  \label{2.14}
\end{equation}
According to this, we can choose to study either one or the other model and
get both solutions at once. We decide to put attention to the model
Hamiltonian (\ref{2.7}), which for a nearest-neighbor potential reads as
\begin{equation}
H=-\mu \sum\limits_{\mathbf{i}}{}\nu (i)+Vd\sum\limits_{\mathbf{i}}{}\nu
(i)\nu ^{\alpha }(i)  \label{2.15}
\end{equation}
where
\begin{equation}
\nu ^{\alpha }(\mathbf{i},t)=\sum\limits_{\mathbf{j}}{}\alpha
_{\mathbf{ij} }\nu (\mathbf{j},t)  \label{2.16}
\end{equation}

The spin-inversion symmetry (\ref{2.5}) of the Ising model (\ref{2.3})
corresponds to the particle-hole symmetry exhibited by the Hamiltonian (\ref
{2.15}).\ In particular, we have that the chemical potential as a function
of $\nu =\left\langle \nu (i)\right\rangle $ scales as
\begin{equation}
\mu (1-\nu )=2dV-\mu (\nu )  \label{2.17}
\end{equation}

\section{Composite operators and equations of motion}

It is immediate to see that the charge density operator $\nu (i)$ satisfies
the equation of motion
\begin{equation}
\text{i}\frac{\partial \nu (i)}{\partial t}=[\nu (i),H]=0  \label{3.0}
\end{equation}
Then, standard methods based on the use of equations of motion and Green's
function (GF) formalism are not immediately applicable. Indeed, it is easy
to check that the causal propagator $\left\langle T[\nu (i)\nu
(j)]\right\rangle $ [$T$ is the chronological operator] and the correlation
function $\left\langle \nu (i)\nu (j)\right\rangle $ assume the form
\begin{equation}
\left\langle T[\nu (i)\nu (j)]\right\rangle =\left\langle \nu (i)\nu
(j)\right\rangle =\frac{1}{N}\sum_{\mathbf{k}}e^{\text{i}\mathbf{k}
\cdot ( \mathbf{R}_{i}-\mathbf{R}_{j})}\Gamma (\mathbf{k})
\label{3.0a}
\end{equation}
where $\Gamma (\mathbf{k})$ is the zero frequency
function\cite{Mancini03} which cannot be calculated by means of the
dynamics\footnote{ Use of the formula $\Gamma
(\mathbf{k})=\frac{1}{2}\lim_{\omega \rightarrow 0}\omega
G^{(+1)}(\mathbf{k},\omega )$, where $G^{(+1)}(\mathbf{k},\omega )$
is the causal propagator defined in terms of fermionic algebra \cite
{Mancini03} would lead just to an identity.}.

Then, in order to solve the Hamiltonian (\ref{2.15}) let us consider the
composite operator
\begin{equation}
\psi _{p}(i)=c(i)[\nu ^{\alpha }(i)]^{p-1}\quad \quad \quad \quad
\{p=1,2\cdots \cdots \}  \label{3.1}
\end{equation}
This field satisfies the equation of motion
\begin{equation}
\text{i}{\frac{\partial }{{\partial t}}}\psi _{p}(i)=[\psi _{p}(i),H]=-\mu
\psi _{p}(i)+2dV\psi _{p+1}(i)  \label{3.2}
\end{equation}
By taking higher-order time derivatives we generate a hierarchy of composite
operators. However, we observe that for $p\geq 1$ the number operator $\nu
(i)=c^{\dag }(i)c(i)$ satisfies the following algebra
\begin{equation}
\lbrack \nu (i)]^{p}=[c^{\dag }(i)c(i)]^{p}=\nu (i)  \label{3.3}
\end{equation}
Therefore, the hierarchy of composite operators (\ref{3.1}) must close for a
certain value of $p$ and we should be able to derive a finite closed set of
eigenoperators of the Hamiltonian. To this purpose, on the basis of (\ref
{3.3}) the following fundamental property of the field $[\nu ^{\alpha
}(i)]^{p}$ can be established
\begin{equation}
\lbrack \nu ^{\alpha }(i)]^{p}=\sum\limits_{m=1}^{2d}{}A_{m}^{(p)}[\nu
^{\alpha }(i)]^{m}  \label{3.4}
\end{equation}
where the coefficients $A_{m}^{(p)}$ satisfy the relation
\begin{equation}
\sum\limits_{m=1}^{2d}{}A_{m}^{(p)}=1  \label{3.5}
\end{equation}
The proof of Eq. (\ref{3.4}) and the explicit expressions of the
coefficients $A_{m}^{(p)}$ are given in Appendix A for the cases $d=1,2,3$.
We now define the composite operator
\begin{equation}
\psi ^{(d)}(i)=\left(
\begin{array}{c}
{{{\psi _{1}(i)}}} \\
{{{\psi _{2}(i)}}} \\
\vdots \\
{{{\psi _{2d+1}(i)}}}
\end{array}
\right) =\left(
\begin{array}{c}
{c(i)} \\
{{{c(i)\nu ^{\alpha }(i)}}} \\
\vdots \\
{{{c(i)[\nu ^{\alpha }(i)]^{2d}}}}
\end{array}
\right)  \label{3.6}
\end{equation}
After (\ref{3.4}), this field is an eigenoperator of the Hamiltonian (\ref
{2.15})
\begin{equation}
\text{i}{\frac{\partial }{{\partial t}}}\psi ^{(d)}(i)=[\psi
^{(d)}(i),H]=\varepsilon ^{(d)}\psi ^{(d)}(i)  \label{3.7}
\end{equation}
where the $(2d+1)\times (2d+1)$ matrix $\varepsilon ^{(d)}$, the
energy matrix, is defined in Appendix B. It is easy to see that the
eigenvalues $ E_{n}^{(d)}$ of the energy matrix are given by
\begin{equation}
E_{n}^{(d)}=-\mu +(n-1)V\quad \quad \quad n=1,2,\cdots,2d+1  \label{3.8}
\end{equation}
The Hamiltonian (\ref{2.15}) has been solved since we know a complete set of
eigenoperators and eigenvalues, and we can proceed to the calculations of
observable quantities. This will be done in the next Sections by using the
Green's function formalism .

\section{Retarded and correlation functions}

We define now the thermal retarded Green's function
\begin{eqnarray}
G^{(d)}(i,j) &=&\left\langle R[\psi ^{(d)}(i)\psi ^{(d)}{}^{\dag
}(j)]\right\rangle  \nonumber \\
&=&\theta (t_{i}-t_{j})\left\langle \{\psi ^{(d)}(i),\psi ^{(d)}{}^{\dag
}(j)\}\right\rangle  \label{4.1}
\end{eqnarray}
where $\left\langle \cdots \right\rangle $ denotes the quantum-statistical
average over the grand canonical ensemble. By introducing the Fourier
transform
\begin{align}
G^{(d)}(i,j)
&={\frac{1}{N}}\sum\limits_{\mathbf{k}}{}{\frac{\text{i}}{{
(2\pi ) }}}  \nonumber \\
&\times \int\limits_{-\infty }^{+\infty }{}d\omega
e^{\text{i}\mathbf{k} \cdot (
\mathbf{R}_{i}-\mathbf{R}_{j})-\text{i}\omega (t_{i}-t_{j})}G^{(d)}(
\mathbf{k},\omega )  \label{4.2}
\end{align}
and by means of the Heisenberg equation (\ref{3.7}) we obtain the equation
\begin{equation}
\lbrack \omega -\varepsilon ^{(d)}]G^{(d)}(\mathbf{k},\omega
)=I^{(d)}( \mathbf{k})  \label{4.3}
\end{equation}
where $I^{(d)}(\mathbf{k})$ is the Fourier transform of the normalization
matrix, defined as
\begin{eqnarray}
I^{(d)}(\mathbf{i,j}) &=&\left\langle \{\psi ^{(d)}(\mathbf{i},t),\psi
^{(d)}{}^{\dag }(\mathbf{j},t)\}\right\rangle  \nonumber \\
&=&{\frac{1}{N}}\sum\limits_{\mathbf{k}}{}e^{\text{i}\mathbf{k}\cdot
( \mathbf{R}_{i}- \mathbf{R}_{j})}I^{(d)}(\mathbf{k})  \label{4.4}
\end{eqnarray}
The solution of Eq. (\ref{4.3}) is
\begin{equation}
G^{(d)}(\mathbf{k},\omega )=\sum\limits_{n=1}^{2d+1}{}{\frac{{\sigma
^{(d,n)}( \mathbf{k})}}{{\omega -E_{n}^{(d)}+}\text{{i}}{\delta }}}
\label{4.5}
\end{equation}
The spectral density matrices $\sigma _{ab}^{(d,n)}(\mathbf{k})$ are
calculated by means of the formula\cite{Mancini03,Mancini04}
\begin{equation}
\sigma _{ab}^{(d,n)}(\mathbf{k})=\Omega _{an}^{(d)}\sum\limits_{c}{}[\Omega
_{nc}^{(d)}]^{-1}I_{cb}^{(d)}(\mathbf{k})  \label{4.6}
\end{equation}
where $\Omega ^{(d)}$is the $(2d+1)\times (2d+1)$ matrix whose columns are
the eigenvectors of the matrix $\varepsilon ^{(d)}$. The explicit
expressions of $\Omega ^{(d)}$are given in Appendix B. The spectral density
matrices $\sigma ^{(d,n)}(\mathbf{k})$ satisfy the sum rule
\begin{equation}
\sum\limits_{n=1}^{2d+1}{}\,[E_{n}^{(d)}]^{p}\sigma ^{(d,n)}(\mathbf{k}
)=M^{(d,p)}(\mathbf{k})  \label{4.7}
\end{equation}
where $M^{(d,p)}(\mathbf{k})$ are the spectral moments defined as
\begin{equation}
M^{(d,p)}(\mathbf{k})=F.T.\left\langle \{\left( {i\partial /\partial
t} \right) ^{p}\psi ^{(d)}(\mathbf{i},t),\psi ^{(d)}{}^{\dag
}(\mathbf{j} ,t)\}\right\rangle  \label{4.8}
\end{equation}
$F.T.$ stays for the Fourier transform. It is a consequence of the
theorem proved in Ref.~\onlinecite{Mancini98} [see also pag.~572 in
Ref.~\onlinecite{Mancini03}] that the spectral density matrices, for
$d=1,2,3$, satisfy the sum rule (\ref{4.7}). The explicit
expressions of $ I^{(d)}( \mathbf{k})$ and $\sigma
^{(d,n)}(\mathbf{k})$ are given in Appendices C and D, respectively,
for the cases $d=1,2,3$. The correlation function
\begin{equation}
C^{(d)}(i,j)=\left\langle \psi ^{(d)}(i)\psi ^{(d)}{}^{\dag }(j)\right\rangle
\label{4.8a}
\end{equation}
can be immediately calculated from (\ref{4.5}) by using the spectral theorem
and one obtains
\begin{align}
&{C^{(d)}(i,j)}  \nonumber \\
&={{\frac{1}{N}}\sum\limits_{\mathbf{k}}{}{\frac{1}{{(2\pi )}}}
\int\limits_{-\infty }^{+\infty }{}d\omega
{e}^{\text{{i}}(\mathbf{R}_{i}- \mathbf{R }_{j})-\text{i}\omega
(t_{i}-t_{j})}C^{(d)}(\mathbf{k},\omega )} \label{4.9}
\end{align}
\begin{equation}
{C^{(d)}(\mathbf{k},\omega )=\pi \sum\limits_{n=1}^{2d+1}{}\delta [\omega
-E_{n}^{(d)}]T_{n}^{(d)}\sigma ^{(d,n)}(\mathbf{k})}  \label{4.9a}
\end{equation}
with
\begin{equation}
T_{n}^{(d)}=1+\tanh \left( {{\frac{{E_{n}^{(d)}}}{{2}k{_{B}T}}}}\right)
\label{4.10}
\end{equation}
Equations (\ref{4.5}) and (\ref{4.9a}) are an exact solution of the
model Hamiltonian (\ref{2.15}). One is able to obtain an exact
solution as the composite operators $\psi _{p}(i)=c(i)[\nu ^{\alpha
}(i)]^{p-1}$ constitute a closed set of eigenoperators of the
Hamiltonian. However, as stressed in Ref.~\onlinecite{Mancini03},
the knowledge of the GF is not fully achieved yet. The algebra of
the field $\psi ^{(d)}(i)$ is not canonical: as a consequence, the
normalization matrix $I^{(d)}(\mathbf{k})$ in the equation (
\ref{4.3}) contains some unknown static correlation functions,
correlators (see Appendix C for explicit calculations), that have to
be self-consistently calculated. According to the scheme of
calculations proposed by the composite operator method \cite
{Mancini03,Mancini03a,Mancini04}(COM), one way of calculating these
unknown correlators is by specifying the representation where the GF
are realized. The knowledge of the Hamiltonian and of the
operatorial algebra is not sufficient to completely determine the
GF. The GF refer to a specific representation (i.e., to a specific
choice of the Hilbert space) and this information must be supplied
to the equations of motion that alone are not sufficient to
completely determine the GF. Usually, the use of composite operators
leads to an enlargement of the Hilbert space by the inclusion of
some unphysical states. Since the GF depend on the unknown
correlators, it is clear that the value of these parameters and the
representation are intimately related. The procedure is the
following. We set up some requirements on the representation and
determine the correlators in order that these conditions be
satisfied. From the algebra it is possible to derive several
relations among the operators. We will call algebra constraints (AC)
all possible relations among the operators dictated by the algebra.
This set of relations valid at microscopic level must be satisfied
also at macroscopic level, when expectations values are considered.
Use of these considerations leads to some self-consistent equations
which will be used to fix the unknown correlators appearing in the
normalization matrix. An immediate set of rules is given by the
equation
\begin{equation}
\left\langle \psi ^{(d)}(i)\psi ^{(d)\mathbf{\dag }}(i)\right\rangle
={\frac{ 1}{ N}}\sum\limits_{\mathbf{k}}{}{\frac{1}{{2\pi
}}}\int\limits_{-\infty }^{+\infty }{}d\omega
\,C^{(d)}(\mathbf{k},\omega )  \label{4.11}
\end{equation}
where the l.h.s. is fixed by the AC and the boundary conditions compatible
with the phase under investigation, while in the r.h.s. the correlation
function $C(\mathbf{k},\omega )$ is computed by means of equation of motion
[cfr. Eq. (\ref{4.9a})].

Another important set of AC can be derived by observing that there exist
some operators, $O$, which project out of the Hamiltonian a reduced part
\begin{equation}
OH=OH_{0}  \label{4.12}
\end{equation}
When $H_{0}$ and $H_{I}=H-H_{0}$ commute, the quantum statistical average of
the operator O over the complete Hamiltonian $H$ must coincide with the
average over the reduced Hamiltonian $H_{0}$
\begin{equation}
Tr\{Oe^{-\beta H}\}=Tr\{Oe^{-\beta H_{0}}\}  \label{4.13}
\end{equation}

Another relation is the requirement of time translational invariance which
leads to the condition that the spectral moments, defined by Eq. (\ref{4.8}
), must satisfy the following relation
\begin{equation}
M_{ab}^{(d,p)}(\mathbf{k})=[M_{ba}^{(d,p)}(\mathbf{k})]^{*}  \label{4.14}
\end{equation}
It can be shown that if (\ref{4.14}) is violated, then states with a
negative norm appear in the Hilbert space. Of course the above rules are not
exhaustive and more conditions might be needed.

According to the calculations given in appendices C and D, the GF and the
correlation functions depend on the following parameters: external
parameters $(\mu ,T,V)$, internal parameters $(C_{1,1}^{(d)\alpha
},C_{1,2}^{(d)\alpha },\cdots C_{1,2d}^{(d)\alpha })$, and $(\kappa
^{(1)},\kappa ^{(1)},\cdots \kappa ^{(2d)})$, defined as
\begin{equation}
C_{\mu ,\nu }^{(d)\alpha }=\left\langle \psi _{\mu }^{(d)\alpha }(i)\psi
_{\nu }^{(d)\dag }(i)\right\rangle  \label{4.15}
\end{equation}
\begin{equation}
\kappa ^{(p)}=\left\langle [v^{\alpha }(i)]^{p}\right\rangle  \label{4.15a}
\end{equation}
The parameters $C_{\mu ,\nu }^{(d)\alpha }$ are determined by means of their
own definitions (\ref{4.15}), where the r.h.s. is calculated by means of (
\ref{4.9})-(\ref{4.9a}). This equation gives
\begin{equation}
C^{(d)\alpha }={\frac{1}{2}}\sum\limits_{n=1}^{2d+1}{}T_{n}^{(d)}{\frac{1}{N}
}\sum\limits_{\mathbf{k}}{}\alpha (\mathbf{k})\sigma ^{(d,n)}(\mathbf{k})
\label{4.16}
\end{equation}
 From the results given in the Appendices C and D, we see that the spectral
density matrices have the form
\begin{equation}
\sigma ^{(d,n)}(\mathbf{k})=\Lambda _{0}^{(d,n)}+\alpha (\mathbf{k})\Lambda
_{1}^{(d,n)}  \label{4.17}
\end{equation}
where the matrices $\Lambda _{0}$ and $\Lambda _{1}$ do not depend on
momentum $\mathbf{k}$. Putting (\ref{4.17}) into (\ref{4.16}) we obtain
\begin{equation}
C^{(d)\alpha }={\frac{1}{{4d}}}\sum\limits_{n=1}^{2d+1}{}T_{n}^{(d)}\Lambda
_{1}^{(d,n)}  \label{4.18}
\end{equation}
Calculations given in the Appendices C and D show that the matrices
$\Lambda _{1}^{(d,n)}$ are linear combinations of the matrix
elements $ C_{1,p}^{(d)\alpha }$. Then, Eq. (\ref{4.18}) gives a
system of homogeneous linear equations. The determinant of this
system is only function of the external parameters $\mu ,T,V$. This
function will vanish only if there is a particular relation among
these parameters. Since these parameters are independent variables
the only solution is that all the matrix elements must vanish
\begin{equation}
C_{1,p}^{(d)\alpha }=0  \label{4.19}
\end{equation}
The matrices $\Lambda _{1}^{(d,n)}$ are zero and the correlation
function $ C^{(d)}(\mathbf{k},\omega )$ does not depend on momentum,
as we expected. In the coordinate space the CF takes the expression
\begin{equation}
C^{(d)}(i,j)=\delta _{\mathbf{ij}}{\frac{1}{2}}\sum
\limits_{n=1}^{2d+1}{}T_{n} \Lambda _{0}^{(d,n)}e^{-\text{i}
E_{n}^{(d)}(t_{i}-t_{j})}  \label{4.20}
\end{equation}
The correlation function depends on $2d$ internal parameters: $\kappa
^{(1)},\cdots ,\kappa ^{(2d)}$. In order to determine these parameters, we
use the Pauli condition (\ref{4.11}) which gives the self-consistent
equations
\begin{equation}
\kappa ^{(p)}-\lambda ^{(p)}=C_{1,p+1}^{(d)}\quad \quad (p=0,1,\cdots 2d)
\label{4.21}
\end{equation}
where $C_{1,p+1}^{(d)}=\left\langle \psi _{1}^{(d)}(i)\psi _{p}^{(d)\dag
}(i)\right\rangle $ is calculated by means of (\ref{4.20}). New correlation
functions
\begin{equation}
\lambda ^{(p)}=\left\langle \nu (i)[\nu ^{\alpha }(i)]^{p}\right\rangle
\label{4.22}
\end{equation}
appear and the set of self-consistent equations (\ref{4.21}) is not
sufficient to determine all unknown parameters. One needs more conditions.
In the case of one-dimensional systems these extra conditions can be
obtained by using the property (\ref{4.13}).

\section{Charge correlations functions}

In Sections 3 and 4, we have solved the problem of the Ising model in terms
of a set of local parameters, defined by (\ref{4.15a}) and (\ref{4.22}). In
this Section, we want to show how we can calculate non-local correlation
functions. Let us define the causal Green's function (for simplicity in this
Section we drop the superindex $(d)$)
\begin{eqnarray}
{F^{C}(i,l,j)} &=&\left\langle {T[\psi (i)\psi ^{\dag }(l)]\nu (j)}
\right\rangle  \nonumber \\
&=&{\theta (t_{i}-t_{l})\left\langle {\psi (i)\psi ^{\dag }(l)\nu (j)}
\right\rangle }  \nonumber \\
&&{-\theta (t_{l}-t_{i})}\left\langle {\psi ^{\dag }(l)\psi (i)\nu (j)}
\right\rangle  \label{6.1a}
\end{eqnarray}
the retarded and advanced functions
\begin{eqnarray}
F^{R,A}(i,l,j) &=&\left\langle R,A[\psi (i)\psi ^{\dag }(l)]\nu
(j)\right\rangle  \nonumber \\
&=&\pm \theta [\pm (t_{i}-t_{l})]{\left\langle \{\psi (i),\psi ^{\dag
}(l)\}\nu (j)\right\rangle }  \label{6.2}
\end{eqnarray}
the correlation functions
\begin{equation}
\begin{array}{l}
{D^{\psi \psi ^{\dag }}(i,l,j)=\left\langle {\psi (i)\psi ^{\dag }(l)\nu (j)}
\right\rangle } \\
{D^{\psi ^{\dag }\psi }(i,l,j)=\left\langle {\psi ^{\dag }(l)\psi (i)\nu (j)}
\right\rangle }
\end{array}
\label{6.3}
\end{equation}
where $\psi (i)$ is the composite field defined in (\ref{3.6}) and we used
the fact the field operator $\nu (j)$ does not depend on time. The Fourier
transforms of these quantities read as
\begin{equation}
\begin{array}{l}
{F^{Q}(i,l,j)={\frac{\text{i}}{{2\pi }}}\int {}d\omega e^{-\text{i}\omega
(t_{i}-t_{l})}F(\mathbf{i,l,j};\omega )} \\
{D^{\psi \psi ^{\dag }}(i,l,j)={\frac{1}{{2\pi }}}\int {}d\omega e^{-\text{i}
\omega (t_{i}-t_{l})}D^{\psi \psi ^{\dag }}(\mathbf{i,l,j};\omega )}
\end{array}
\label{6.4}
\end{equation}
where $Q=C,R,A.$ By means of the equation of motion (\ref{3.7}) we have
\begin{equation}
(\omega -\varepsilon )F^{Q}{(\mathbf{i,l,j};\omega )}=J{(\mathbf{i,l,j})}
\label{6.5}
\end{equation}
\begin{equation}
\begin{array}{l}
{(\omega -\varepsilon )D^{\psi \psi ^{\dag }}(\mathbf{i,l,j};\omega )=0} \\
{(\omega -\varepsilon )D^{\psi ^{\dag }\psi }(\mathbf{i,l,j};\omega )=0}
\end{array}
\label{6.6}
\end{equation}
where the matrix $J{(\mathbf{i,l,j})}$ is defined as
\begin{equation}
J{(\mathbf{i,l,j})}=\left\langle \{\psi (\mathbf{i},t),\psi ^{\dag
}(\mathbf{ l},t)\}n( \mathbf{j})\right\rangle  \label{6.7}
\end{equation}

The most general solution of Eq. (\ref{6.5}) is
\begin{eqnarray}
F^{Q}{(\mathbf{i,l,j};\omega )} &=&\sum\limits_{n=1}^{2d+1}{}[P{\frac{{\tau
^{(n)}(\mathbf{i,l,j})}}{{\omega -E_{n}}}}  \nonumber \\
&&-\text{i}\pi \delta (\omega -E_{n})g^{Q(n)}{(\mathbf{i,l,j})}]  \label{6.8}
\end{eqnarray}
where
\begin{equation}
\tau _{ab}^{(n)}{(\mathbf{i,l,j})}=\Omega
_{an}\sum\limits_{c=1}^{2d+1}{}\Omega _{nc}^{-1}J_{cb}{(\mathbf{i,l,j})}
\label{6.9}
\end{equation}
while the function $g^{Q(n)}{(\mathbf{i,l,j})}$ must be determined. $P$
denotes the principal value. By recalling the retarded and advanced nature
of $F^{R,A}(i,l,j)$, it is immediate to see that
\begin{equation}
g^{R(n)}{(\mathbf{i,l,j})}=-g^{A(n)}{(\mathbf{i,l,j})}=\tau
^{(n)}{(\mathbf{ i,l,j})}  \label{6.10}
\end{equation}
Therefore
\begin{equation}
F^{R,A}{(\mathbf{i,l,j};\omega )}=\sum\limits_{n=1}^{2d+1}\frac{{\tau
^{(n)}( \mathbf{i,l,j})}}{{\omega -E_{n}\pm }\text{{i}}{\delta }}
\label{6.11}
\end{equation}
The solution of (\ref{6.6}) is
\begin{equation}
\begin{array}{l}
D^{\psi \psi ^{\dag }}{(\mathbf{i,l,j};\omega
)=\sum\limits_{n=1}^{2d+1}{}\delta (\omega -E_{n})d^{\psi \psi ^{\dag }(n)}(
\mathbf{i,l,j})} \\
D^{\psi ^{\dag }\psi }{(\mathbf{i,l,j};\omega
)=\sum\limits_{n=1}^{2d+1}{}\delta (\omega -E_{n})d^{\psi ^{\dag }\psi (n)}(
\mathbf{i,l,j})}
\end{array}
\label{6.12}
\end{equation}
where the matrices ${d^{\psi \psi ^{\dag }(n)}(\mathbf{i,l,j})}$ and
${ d^{\psi ^{\dag }\psi (n)}(\mathbf{i,l,j})}$ have to be
determined. From the definitions (\ref{6.1a})-(\ref{6.3}) we can
derive the following exact relations
\begin{equation}
\begin{array}{l}
{F^{R}(i,l,j)+F^{A}(i,l,j)=2F^{C}(i,l,j)-\left\langle {[\psi (i),\psi ^{\dag
}(l)]\nu (j)}\right\rangle } \\
{F^{R}(i,l,j)-F^{A}(i,l,j)=\left\langle {\{\psi (i),\psi ^{\dag }(l)\}\nu
(j) }\right\rangle }
\end{array}
\label{6.13}
\end{equation}
A relation between the two correlation functions $D^{\psi \psi ^{\dag
}}(i,l,j)$ and $D^{\psi ^{\dag }\psi }(i,l,j)$ can be established by means
of trace properties. Indeed, it is straightforward to derive a KMS-like
relation
\begin{eqnarray}
\left\langle \psi ^{\dag }(l)\psi (i)\nu (j)\right\rangle &=&\left\langle
\psi (i,t_{i}-i\beta )\psi ^{\dag }(l)\nu (j)\right\rangle  \nonumber \\
&&+\delta _{\mathbf{lj}}\left\langle \psi (i,t_{i}-i\beta )\psi ^{\dag
}(l)\right\rangle  \label{6.14}
\end{eqnarray}
By recalling the definitions (\ref{6.3}), this last equations can be written
as
\begin{equation}
D^{\psi ^{\dag }\psi }{(\mathbf{i,l,j};\omega )}=e^{-\beta \omega
}[D^{\psi \psi ^{\dag }}{(\mathbf{i,l,j};\omega )}+\delta
_{\mathbf{lj}}C(\mathbf{i,l} ;\omega )]  \label{6.15}
\end{equation}
where $C(\mathbf{i,l};\omega )$ is the fermionic correlation function [see
Eq. ( \ref{4.9})-(\ref{4.9a})]. Therefore, the anticommutator $\left\langle
\{\psi (i),\psi ^{\dag }(l)\}\nu (j)\right\rangle $ in (\ref{6.13}) can be
expressed in terms of the correlation functions as
\begin{equation}
\begin{array}{l}
\left\langle \{\psi (i),\psi ^{\dag }(l)\}\nu (j)\right\rangle ={\frac{1}{{2
\pi }}}\int {}d\omega e^{-i\omega (t_{i}-t_{l})} \\
\times [(1+e^{-\beta \omega })D^{\psi \psi ^{\dag
}}{(\mathbf{i,l,j};\omega ) } +\delta _{\mathbf{lj}}e^{-\beta \omega
}C(\mathbf{i,l};\omega )]
\end{array}
\label{6.17}
\end{equation}
Analogous expression holds for the commutator. By means of (\ref{6.17}) and
by recalling that [see Eqs. (\ref{4.9})-(\ref{4.9a})]
\begin{equation}
\begin{array}{l}
{(\omega -\varepsilon )C(\mathbf{i,l};\omega )=0} \\
{C(\mathbf{i,l};\omega )=\sum\limits_{n=1}^{2d+1}{}\delta (\omega
-E_{n})c^{(n)}(\mathbf{i,l})}
\end{array}
\label{6.18}
\end{equation}
we find that equations (\ref{6.13}) have the following form
\begin{equation}
\begin{array}{l}
{\sum\limits_{n=1}^{2d+1}{}\delta (\omega
-E_{n})\{g^{C(n)}(\mathbf{i,l,j})-{
\ \frac{1}{{2\pi }}}[(1-e^{-\beta \omega })} \\
\times {d^{\psi \psi ^{\dag }(n)}(\mathbf{i,l,j})-\delta
_{\mathbf{lj} }e^{-\beta \omega }c^{(n)}(\mathbf{i,l})]\}=0}
\end{array}
\label{6.19}
\end{equation}
\begin{equation}
\begin{array}{l}
{\sum\limits_{n=1}^{2d+1}{}\delta (\omega -E_{n})\{\tau
^{(n)}(\mathbf{i,l,j}
)-{\ \frac{1}{{2\pi }}}[(1+e^{-\beta \omega })} \\
\times {d^{\psi \psi ^{\dag }(n)}(\mathbf{i,l,j})+\delta
_{\mathbf{lj} }e^{-\beta \omega }c^{(n)}(\mathbf{i,l})]\}=0}
\end{array}
\label{6.20}
\end{equation}
By recalling that $E_{n}^{(d)}=-\mu +(n-1)V$, the solution of (\ref{6.19})
and (\ref{6.20},) is:
\begin{eqnarray}
{d^{\psi \psi \dag (n)}(\mathbf{i,l,j})} &=&{{\frac{{2\pi }}{{1+e^{-\beta
E_{n}} }}}\tau ^{(n)}(\mathbf{i,l,j})}  \nonumber \\
&&{-{\frac{{\delta _{\mathbf{lj}}e^{-\beta E_{n}}}}{{1+e^{-\beta E_{n}}}}}
c^{(n)}(\mathbf{i,l})}  \label{6.20a}
\end{eqnarray}
\begin{eqnarray}
{g^{C(n)}(\mathbf{i,l,j})} &=&{{\frac{{1-e^{-\beta E_{n}}}}{{1+e^{-\beta
E_{n}}} }}\tau ^{(n)}(\mathbf{i,l,j})}  \nonumber \\
&&{-{\frac{{\delta _{\mathbf{lj}}}}{{2\pi }}}{\frac{{2e^{-\beta E_{n}}}}{{\
1+e^{-\beta E_{n}}}}}c^{(n)}(\mathbf{i,l})}  \label{6.20b}
\end{eqnarray}
By putting (\ref{6.20a}) and (\ref{6.20b}) into (\ref{6.8}) and (\ref{6.12})
we have
\begin{eqnarray}
D^{\psi \psi ^{\dag }}(i,l,j) &=&\sum\limits_{n=1}^{2d+1}{}{\frac{{e^{-\text{
i}E_{n}(t_{i}-t_{l})}}}{{1+e^{-\beta E_{n}}}}}[\tau ^{(n)}(\mathbf{i,l,j})
\nonumber \\
&&-{\frac{1}{{2\pi }}\delta _{\mathbf{lj}}}e^{-\beta
E_{n}}{c^{(n)}(\mathbf{ i,l})}]  \label{6.21}
\end{eqnarray}
\begin{eqnarray}
D^{\psi ^{\dag }\psi }(i,l,j) &=&\sum\limits_{n=1}^{2d+1}{}{\frac{{e^{-\text{
i}E_{n}(t_{i}-t_{l})}e^{-\beta E_{n}}}}{{1+e^{-\beta E_{n}}}}}[\tau ^{(n)}(
\mathbf{i,l,j})  \nonumber \\
&&+{\frac{1}{{2\pi }}}\delta _{\mathbf{lj}}{c^{(n)}(\mathbf{i,l})}]
\label{6.22}
\end{eqnarray}
\begin{eqnarray}
{F^{C}(i,l,j)} &=&{{\frac{\text{i}}{{2\pi }}}\int {}d\omega
e^{-\text{i} \omega
(t_{i}-t_{l})}\sum\limits_{n=1}^{2d+1}{}{\frac{{\tau ^{(n)}(\mathbf{
i,l,j })}}{{\ 1+e^{-\beta E_{n}}}}}}  \nonumber \\
&&\times {\left[ {{\frac{1}{{\omega -E_{n}+}\text{{i}}{\delta }}}+{\frac{{\
e^{-\beta E_{n}}}}{{\omega -E_{n}-}\text{{i}}{\delta }}}}\right] }  \nonumber
\\
&&{-{\frac{\text{i}}{{2\pi }}}\delta _{\mathbf{l,j}}\int {}d\omega
e^{-i\omega
(t_{i}-t_{l})}\sum\limits_{n=1}^{2d+1}{}{\frac{{c^{(n)}(\mathbf{
i,l})e^{-\beta E_{n}}}}{{1+e^{-\beta E_{n}}}}}}  \nonumber \\
&&{\left[ {{\frac{1}{{\omega -E_{n}+}\text{{i}}{\delta }}}-{\frac{1}{{\omega
-E_{n}-}\text{{i}}{\delta }}}}\right] }  \label{6.23}
\end{eqnarray}
 From the study of the fermionic sector we have
\begin{equation}
c^{(n)}{(\mathbf{i,l})}={\frac{{2\pi }}{{1+e^{-\beta E_{n}}}}}\delta
_{ \mathbf{il} }\sigma ^{(n)}  \label{6.24}
\end{equation}
where $\sigma ^{(n)}$are the spectral functions given in Appendix D. Then,
at equal time (\ref{6.21}) becomes
\begin{align}
&D^{\psi \psi ^{\dag }}{(\mathbf{i,l,j})}
=\sum\limits_{n=1}^{2d+1}{}{\frac{1
}{{1+e^{-\beta E_{n}}}}}  \nonumber \\
&\times [\tau ^{(n)}{(\mathbf{i,l,j})}-\delta _{\mathbf{il}}\delta
_{\mathbf{ lj}}{\ \frac{1}{{1+e^{\beta E_{n}}}}}\sigma ^{(n)}]
\label{6.25}
\end{align}

The system (\ref{6.25}) gives a system of linear equations for the
quantities $D^{\psi \psi ^{\dag }}{(\mathbf{i,l,j})}$. Since the
inhomogeneous terms in this system are proportional to $\delta
_{\mathbf{il} } $, it is clear that $D^{\psi \psi ^{\dag
}}{(\mathbf{i,l,j})}\propto \delta _{\mathbf{il}}$. Then, we will
take $\mathbf{i}=\mathbf{l}$ and we write
\begin{equation}
D(\mathbf{i,j})=\left\langle \psi (\mathbf{i})\psi ^{\dag
}(\mathbf{i})n( \mathbf{j} )\right\rangle  \label{6.26}
\end{equation}
 From (\ref{6.25}) we have the system of equations
\begin{equation}
D(\mathbf{i,j})=\sum\limits_{n=1}^{2d+1}{}{\frac{1}{{1+e^{-\beta
E_{n}}}}} [\tau ^{(n)}(\mathbf{i,j})-\delta
_{\mathbf{ij}}{\frac{1}{{1+e^{\beta E_{n}}} }}\sigma ^{(n)}]
\label{6.27}
\end{equation}
where
\begin{equation}
\begin{array}{l}
{\tau _{ab}^{(n)}(\mathbf{i,j})=\Omega _{an}\sum\limits_{c=1}^{2d+1}{}\Omega
_{nc}^{-1}J_{cb}(\mathbf{i,j})} \\
{J(\mathbf{i,j})=}\left\langle \psi (\mathbf{i})\psi ^{\dag }(\mathbf{i})\nu
(\mathbf{j} )\right\rangle
\end{array}
\label{6.28}
\end{equation}
 From its own definition (\ref{6.26}) and by using the recurrence relation (
\ref{3.4}), the matrix $D{(\mathbf{i,j})}$ has the following structure.

\textbf{(i) One dimension}
\begin{equation}
D^{(1)}{(\mathbf{i,j})=}\left(
\begin{array}{ccc}
D_{1,1} & D_{1,2} & D_{1,3} \\
D_{1,2} & D_{1,3} & D_{2,3} \\
D_{1,3} & D_{2,3} & D_{3,3}
\end{array}
\right)  \label{6.29a}
\end{equation}

\begin{equation}
\begin{array}{l}
{{D_{1,p}(\mathbf{i,j})=K^{(p-1)}(\mathbf{i,j})-\Lambda
^{(p-1)}(\mathbf{i,j}
)\qquad p=1,2,3}} \\
{{D_{p,3}(\mathbf{i,j})=\sum\limits_{m=1}^{2}{}A_{m}^{(p+1)}D_{1,m+1}(
\mathbf{i,j} )\qquad \quad \,\,p=2,3}}
\end{array}
\label{6.30}
\end{equation}

\textbf{(ii) Two dimensions}
\begin{equation}
D^{(2)}{(\mathbf{i,j})=}\left(
\begin{array}{ccccc}
D_{1,1} & D_{1,2} & D_{1,3} & D_{1,4} & D_{1,5} \\
D_{1,2} & D_{1,3} & D_{1,4} & D_{1,5} & D_{2,5} \\
D_{1,3} & D_{1,4} & D_{1,5} & D_{2,5} & D_{3,5} \\
D_{1,4} & D_{1,5} & D_{2,5} & D_{3,5} & D_{4,5} \\
D_{1,5} & D_{2,5} & D_{3,5} & D_{4,5} & D_{5,5}
\end{array}
\right)  \label{6.31}
\end{equation}
\begin{equation}
\begin{array}{l}
{{D_{1,p}(\mathbf{i,j})=K^{(p-1)}(\mathbf{i,j})-\Lambda
^{(p-1)}(\mathbf{i,j}
)\quad p=1,2,\cdots ,5}} \\
{{D_{p,5}(\mathbf{i,j})=\sum\limits_{m=1}^{4}{}A_{m}^{(p+3)}D_{1,m+1}(
\mathbf{i,j} )\qquad \,\,p=2,3,\cdots ,5}}
\end{array}
\label{6.32}
\end{equation}

\textbf{(iii) Three dimensions}
\begin{equation}
D^{(3)}{(\mathbf{i,j})=}\left(
\begin{array}{ccccc}
D_{1,1} & D_{1,2} & \cdots & D_{1,6} & D_{1,7} \\
D_{1,2} & D_{1,3} & \cdots & D_{1,5} & D_{2,7} \\
\vdots & \vdots & \vdots & \vdots & \vdots \\
D_{1,6} & D_{1,7} & \cdots & D_{5,7} & D_{6,7} \\
D_{1,7} & D_{2,7} & \cdots & D_{6,7} & D_{7,7}
\end{array}
\right)  \label{6.33}
\end{equation}
\begin{equation}
\begin{array}{l}
{{D_{1,p}(\mathbf{i,j})=K^{(p-1)}(\mathbf{i,j})-\Lambda
^{(p-1)}(\mathbf{i,j}
)\quad p=1,2,\cdots ,7}} \\
{{D_{p,7}(\mathbf{i,j})=\sum\limits_{m=1}^{6}{}A_{m}^{(p+5)}D_{1,m+1}(
\mathbf{i,j} )\qquad \,\,p=2,3,\cdots ,7}}
\end{array}
\label{6.34}
\end{equation}

With the definitions
\begin{equation}
\begin{array}{l}
{K^{(p)}(\mathbf{i,j})=<[\nu ^{\alpha }(\mathbf{i})]^{p}\nu (\mathbf{j})>}
\\
{\Lambda ^{(p)}(\mathbf{i,j})=<\nu (\mathbf{i})[\nu ^{\alpha
}(\mathbf{i} )]^{p}\nu ( \mathbf{j})>}
\end{array}
\label{6.35}
\end{equation}
Then, we only need to calculate the matrix elements
$D_{1,p}(\mathbf{i,j} )\quad (p=1,2,\cdots 2d+1)$. The matrix
$J{(\mathbf{i,j})}$ can be obtained from the normalization matrix
$I{(\mathbf{i,j})}=\left\langle \{\psi ( \mathbf{i},t),\psi ^{\dag
}(\mathbf{j},t)\}\right\rangle $, calculated in Appendix C, by means
of the following substitution
\begin{equation}
\kappa ^{(p)}\quad \to \quad K^{(p)}(\mathbf{i,j})  \label{6.36}
\end{equation}
Then, the matrices $\tau ^{(n)}{(\mathbf{i,j})}$ have the same expressions
of the spectral matrices $\sigma ^{(n)}$ when the following substitution
\begin{equation}
I_{ab}\quad \to \quad J_{ab}{(\mathbf{i,j})}  \label{6.37}
\end{equation}
is made. It can be seen that for $\mathbf{j}=\mathbf{i}$ and
$\mathbf{j}= \mathbf{i} ^{\alpha }$ the system (\ref{6.27}) is
exactly equivalent to the system (\ref{4.21}). Then, it is enough to
consider the case $\mathbf{j}\ne \mathbf{i,i} ^{\alpha }$. In this
case, the system (\ref{6.27}) becomes
\begin{equation}
D{(\mathbf{i,j})}={\frac{1}{2}}\sum\limits_{n=1}^{2d+1}{}T_{n}\tau
^{(n)}{( \mathbf{i,j})}  \label{6.38}
\end{equation}
with $T_{n}$ given by (\ref{4.10}). The system (\ref{6.38}) gives a set of
exact relations among the correlation functions. We might think to solve
this system by induction method, since some of the first correlation
functions can be expressed in terms of the basic parameters $\kappa ^{(p)}$
and $\lambda ^{(p)}$.\ However, when we do this, we immediately see that the
number of equations is not sufficient to determine all the correlation
functions and we need more equations. Once again, this can be done for the
one dimensional system, as we shall see in the next Sections.

\section{Self-consistent equations for one-dimensional systems}

Until now the analysis has been carried on in complete generality for a
cubic lattice of $d$ dimensions. We now consider one-dimensional systems,
and in particular we will study an infinite chain in the homogeneous phase.
For simplicity of notation we shall drop the superindex $(d)$. By means of (
\ref{C6}) and (\ref{D1})-(\ref{D2}) the set of equations (\ref{4.21}) gives
the linear system
\begin{equation}
\begin{array}{l}
T_{1}-2+(2-3T_{1}+4T_{2}-T_{3})\nu \\
+2(T_{1}-2T_{2}+T_{3})\kappa ^{(2)}=0 \\
(2T_{2}-T_{3}-2)\nu -2(T_{2}-T_{3})\kappa ^{(2)}+2\lambda ^{(1)}=0 \\
(T_{2}-T_{3})\nu -(2+T_{2}-2T_{3})\kappa ^{(2)}+2\lambda ^{(2)}=0
\end{array}
\label{5.1}
\end{equation}
where, because of translational invariance, we put
\begin{equation}
\nu =\left\langle \nu (i)\right\rangle =\kappa ^{(1)}  \label{5.2}
\end{equation}
It is immediate to see that for $\mu =V$, the solution of the first equation
in (\ref{5.1}) for $T>0$ is
\begin{equation}
\nu ={\frac{1}{2}\qquad }\text{{for}}{\quad }\mu =V  \label{5.3}
\end{equation}
This is in agreement with the particle-hole symmetry enjoyed by the
model [see (\ref{2.17})]. Recalling (\ref{2.8}) and (\ref{2.11}),
this situation corresponds to the zero magnetization of the Ising
model in absence of external magnetic field. Coming back to general
value of $\mu $, it is clear that Eqs. (\ref{5.1}) are not
sufficient to specify completely the 4 parameters $\nu ,\kappa
^{(2)},\lambda ^{(1)},\lambda ^{(2)}$ and we need another equation.
A fourth equation can be easily obtained by means of the algebra. We
observe that
\begin{equation}
c^{\dag }(i)\nu (i)=0  \label{5.4}
\end{equation}
This relation leads to
\begin{equation}
c^{\dag }(i)e^{-\beta H}=c^{\dag }(i)e^{-\beta H_{0}}  \label{5.5}
\end{equation}
where
\begin{equation}
H_{0}=H-2V\nu (i)\nu ^{\alpha }(i)  \label{5.6}
\end{equation}
By means of the requirement (\ref{4.13}) the correlation function $
C_{1,k}=\left\langle c(i)c^{\dag }(i)[\nu ^{\alpha
}(i)]^{k-1}\right\rangle $ can be written as
\begin{equation}
{\frac{{C_{1,k}}}{{C_{1,1}}}}={\frac{{C_{1,k}^{(0)}}}{{C_{1,1}^{(0)}}}}
\label{5.7}
\end{equation}
where
\begin{equation}
C_{1,k}^{(0)}=\left\langle c(i)c^{\dag }(i)[v^{\alpha
}(i)]^{k-1}\right\rangle _{0}  \label{5.8}
\end{equation}
and $\left\langle \cdots \right\rangle _{0}$denotes the thermal average with
respect to $H_{0}$. Let us define the retarded GF
\begin{eqnarray}
G_{1,k}^{(0)}(t-t^{\prime }) &=&\left\langle R[c(i,t)c^{\dag }(i,t^{\prime
})][v^{\alpha }(i)]^{k-1}\right\rangle _{0}  \nonumber \\
&=&{\frac{\text{i}}{{2\pi }}}\int\limits_{-\infty }^{+\infty
}{}d\omega e^{- \text{i}\omega (t-t)}G_{1,k}^{(0)}(\omega )
\label{5.9a}
\end{eqnarray}
By means of the equation of motion
\begin{equation}
\lbrack c(i),H_{0}]=-\mu c(i)  \label{5.10}
\end{equation}
we have
\begin{equation}
G_{1,k}^{(0)}(\omega )={\frac{\left\langle {[\nu ^{\alpha
}(i)]^{k-1}} \right\rangle {_{0}}}{{\omega +\mu +}\text{{i}}{\delta
}}}  \label{5.11}
\end{equation}
Recalling the relation between retarded and correlation functions, from (\ref
{5.11}) we obtain
\begin{equation}
C_{1,k}^{(0)}={\frac{\left\langle {[\nu ^{\alpha
}(i)]^{k-1}}\right\rangle { _{0}}}{{1+e^{\beta \mu }}}}
\label{5.12}
\end{equation}
By putting this result into (\ref{5.7}) we have
\begin{equation}
C_{1,k}=C_{1,1}\left\langle [\nu ^{\alpha }(i)]^{k-1}\right\rangle _{0}
\label{5.13}
\end{equation}
By noting that $[\nu ^{\alpha }(i)]^{2}$can be expressed as [cfr. (\ref{A4}
)]
\begin{equation}
\lbrack \nu ^{\alpha }(i)]^{2}={\frac{1}{2}}[\nu ^{\alpha }(i)+\nu
(i_{1})\nu (i_{2})]  \label{5.14}
\end{equation}
we obtain from (\ref{5.13}) the relations
\begin{eqnarray}
C_{1,2} &=&C_{1,1}\left\langle [\nu ^{\alpha }(i)]\right\rangle _{0}
\label{5.15} \\
C_{1,3} &=&{\frac{1}{2}}[C_{1,2}+C_{1,1}\left\langle \nu (i_{1})\nu
(i_{2})\right\rangle _{0}]  \label{5.16}
\end{eqnarray}
Now, we observe\cite{Fedro76} that $H_{0}$ describes a system where the
original lattice is divided in two disconnected sublattices (the chains to
the left and to the right of the site $\mathbf{i}$). Then, in $H_{0}$
-representation, the correlation function which relates sites belonging to
different sublattices can be decoupled:
\begin{equation}
\left\langle a(j)b(m)\right\rangle _{0}=\left\langle a(j)\right\rangle
_{0}\left\langle b(m)\right\rangle _{0}  \label{5.17}
\end{equation}
for $j$ and $m$ belonging to different sublattices. By using this property,
invariance of $H_{0}$ under axis reflection and (\ref{5.15}) we can write
\begin{eqnarray}
\left\langle \nu (i_{1})\nu (i_{2})\right\rangle _{0} &=&\left\langle \nu
(i_{1})\right\rangle _{0}\left\langle \nu (i_{2})\right\rangle _{0}
\nonumber \\
&=&[\left\langle \nu ^{\alpha }(i)\right\rangle _{0}]^{2}=\left[
{{\frac{{ C_{1,2}}}{{C_{1,1}}}}}\right] ^{2}  \label{5.18}
\end{eqnarray}
By putting (\ref{5.18}) into (\ref{5.16}), we obtain the following
self-consistent equation among the correlation functions
\begin{equation}
C_{1,3}={\frac{1}{2}}C_{1,2}\left( {1+{\frac{{C_{1,2}}}{{C_{1,1}}}}}\right)
\label{5.19}
\end{equation}
By means of (\ref{4.9}) and (\ref{4.9a}) and the results given in Appendices
C and D, Eq. (\ref{5.19}) takes the expression
\begin{equation}
\begin{array}{l}
(4T_{2}^{2}-{3T_{1}T_{3})\nu }^{2}+[{T_{1}T_{3}-8}\kappa
^{(2)}(T_{2}^{2}-{
T_{1}T_{3}})]\nu \\
+2\kappa ^{(2)}[2\kappa ^{(2)}(T_{2}^{2}-{T_{1}T_{3}})-{T_{1}T_{3}]=0}
\end{array}
\label{5.19a}
\end{equation}
This equation together with equations (\ref{5.1}) gives a system of 4
self-consistent equations for the 4 parameters $\nu ,\kappa ^{(2)},\lambda
^{(1)},\lambda ^{(2)}$ as functions of $\mu ,T,V$. By solving the set of
linear equations (\ref{5.1}) with respect to $\kappa ^{(2)},\lambda
^{(1)},\lambda ^{(2)}$ as functions of $\nu $, we have
\begin{eqnarray}
\kappa ^{(2)} &=&{\frac{{2-T_{1}+\nu (-2+3T_{1}-4T_{2}+T_{3})}}{{\
2(T_{1}-2T_{2}+T_{3})}}}  \label{5.20} \\
\lambda ^{(1)} &=&{\frac{{1}}{{2(T_{1}-2T_{2}+T_{3})}}}\{{\
(2-T_{1})(T_{2}-T_{3})}  \label{5.21} \\
&&{+\nu [T_{1}(2+T_{2}-2T_{3})+T_{2}(T_{3}-6)+4T_{3}]\}}  \nonumber \\
\lambda ^{(2)} &=&{\frac{{1}}{{4(T_{1}-2T_{2}+T_{3})}}}\{{\
(2-T_{1})(2+T_{2}-2T_{3})}  \label{5.22} \\
&&{+\nu [-4-10T_{2}+T_{1}(6+T_{2}-4T_{3})+6T_{3}+3T_{2}T_{3}]\}}  \nonumber
\end{eqnarray}
To calculate the parameter $\nu $ let us put (\ref{5.20}) into (\ref{5.19a})
and solve with respect to $\nu $. We have two roots. One solution
corresponds to an unstable state with negative compressibility and must be
disregarded. By picking up the right root, and by using the relation
\begin{equation}
T_{3}=\frac{2T_{2}^{2}(2-T_{1})}{T_{1}(2-T_{2})^{2}+T_{2}^{2}(2-T_{1})}
\label{5.22a}
\end{equation}
we find
\begin{equation}
\nu =\frac{1}{2}\left[ 1+(1-T_{2})\sqrt{\frac{T_{1}}{
T_{1}-2T_{1}T_{2}+2T_{2}^{2}}}\right]  \label{5.23}
\end{equation}
As shown in Appendix E, the solutions (\ref{5.20})-(\ref{5.23})
exactly correspond to the well-known solution of the 1D Ising model,
obtained by means of the transfer matrix method. We could manipulate
the expression (\ref {5.23}) and the ones for $\kappa ^{(2)},\lambda
^{(1)}$ and $\lambda ^{(2)}$ , obtained by substituting (\ref{5.23})
into (\ref{5.20})-(\ref{5.22}), in order to reproduce the
expressions of the Ising model, given in Appendix E. However, we
prefer to maintain the present expressions as the following
discussion will be more transparent. In Section 3, we have seen that
in the present model, in the one-dimensional case, there are three
energy scales
\begin{equation}
\begin{array}{l}
E_{1}=-\mu \\
E_{2}=-\mu +V \\
E_{3}=-\mu +2V
\end{array}
\label{5.26}
\end{equation}

\begin{figure}[ptb]
\includegraphics[width=8cm]{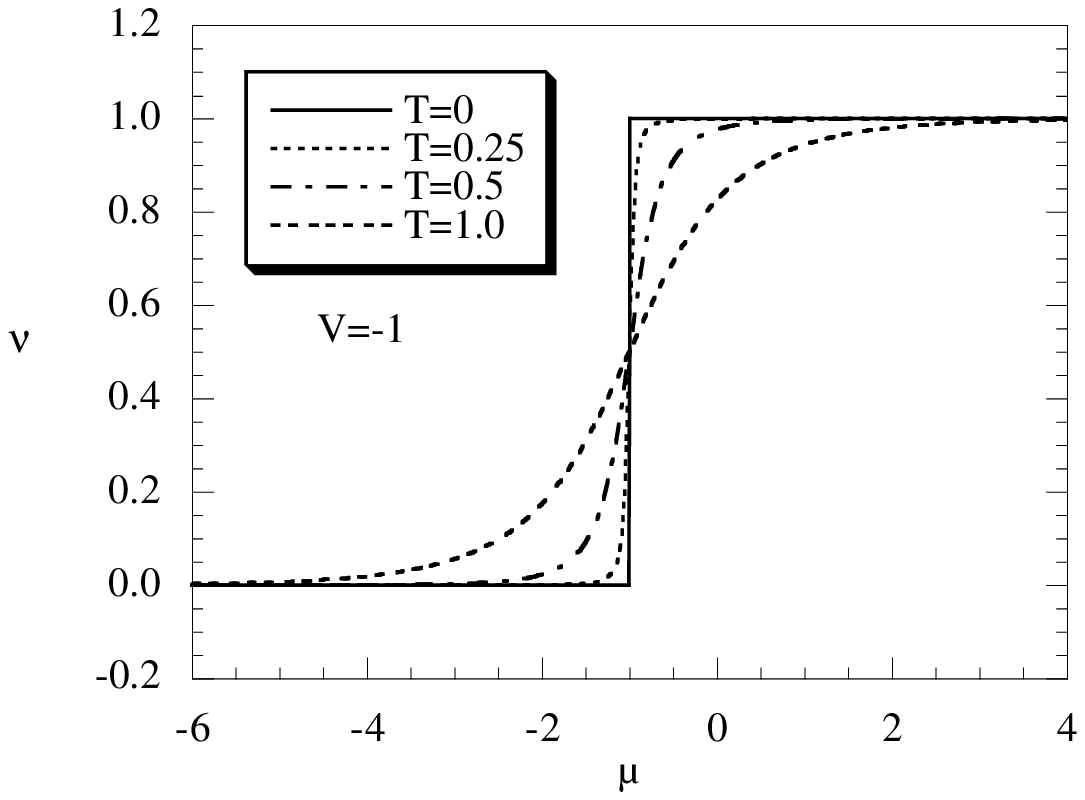} \includegraphics[width=8cm]{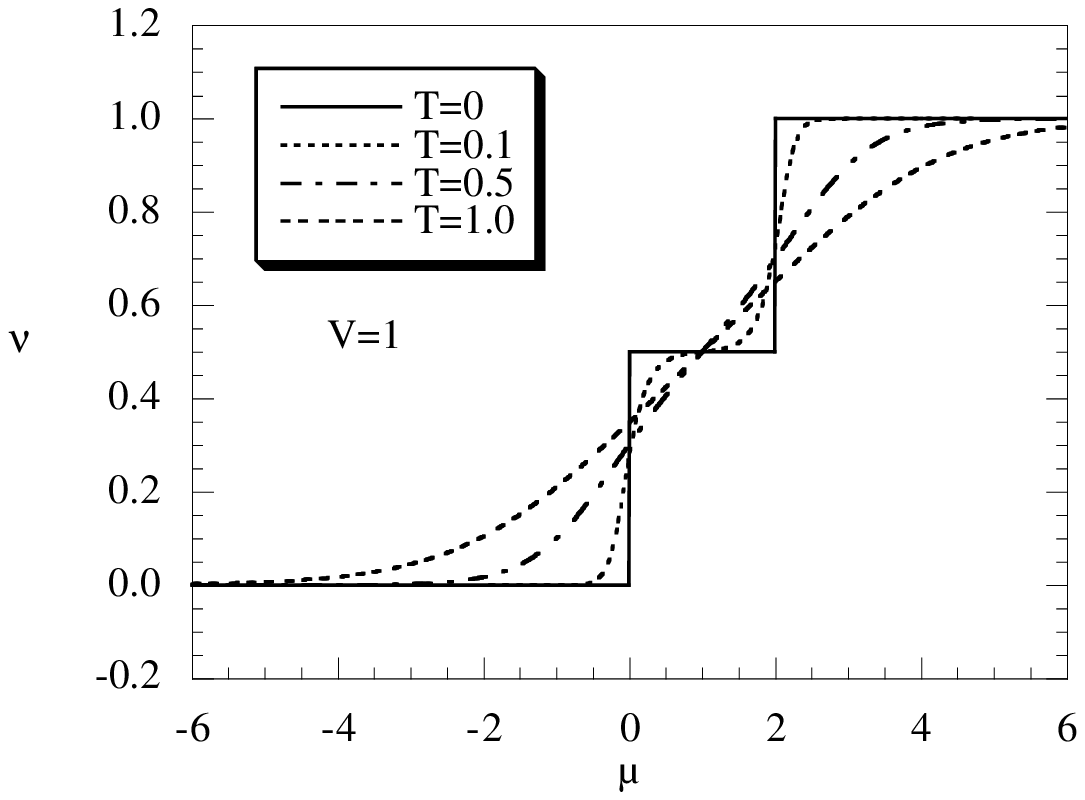}
\caption{The particle density $\nu$ is plotted as a function of the
chemical potential at various temperatures for $V=-1$ (top) and
$V=1$ (bottom).} \label{Fig1}
\end{figure}

At zero temperature, the three functions $T_{1}$, $T_{2}$ and
$T_{3}$ are not analytical functions at the points $\mu =0,$ $\mu
=V,$ and $\mu =2V$, respectively, and we expect that the parameters
$\nu ,\kappa ^{(2)},\lambda ^{(1)},\lambda ^{(2)}$ exhibit some
discontinuous behavior at these points. As shown in Fig.~\ref{Fig1},
in the limit $T\rightarrow 0$ the particle density $\nu $ has a
discontinuity at $\mu =V$ for the case of negative $V$ (i.e.\ $J>0$,
ferromagnetic coupling) and two discontinuities at $\mu =0$ and $\mu
=2V$ for the case of positive $V$ (i.e.\ $J<0$, antiferromagnetic
coupling).\ Here and in the following, we take $\left| V\right| =1$:
all energies are measured in units of $\left| V\right| $. In
particular, the particle density increases by increasing $\mu $ from
zero to one.\ At zero temperature, in the ferromagnetic case $\nu $
is zero for $\mu <-\left| V\right| $ and equal to one for $\mu
>-\left| V\right| $; in the antiferromagnetic case $\nu $ is zero
for $\mu <0$, jumps to $1/2$ and exhibits a plateau, centered at
$\mu =V$, in the region $0<\mu <2V$, jumps to the value $1$ for $\mu
>2V$. The parameter $\kappa ^{(2)}$ has a behavior similar to $\nu
$.

\begin{figure}[ptb]
\includegraphics[width=8cm]{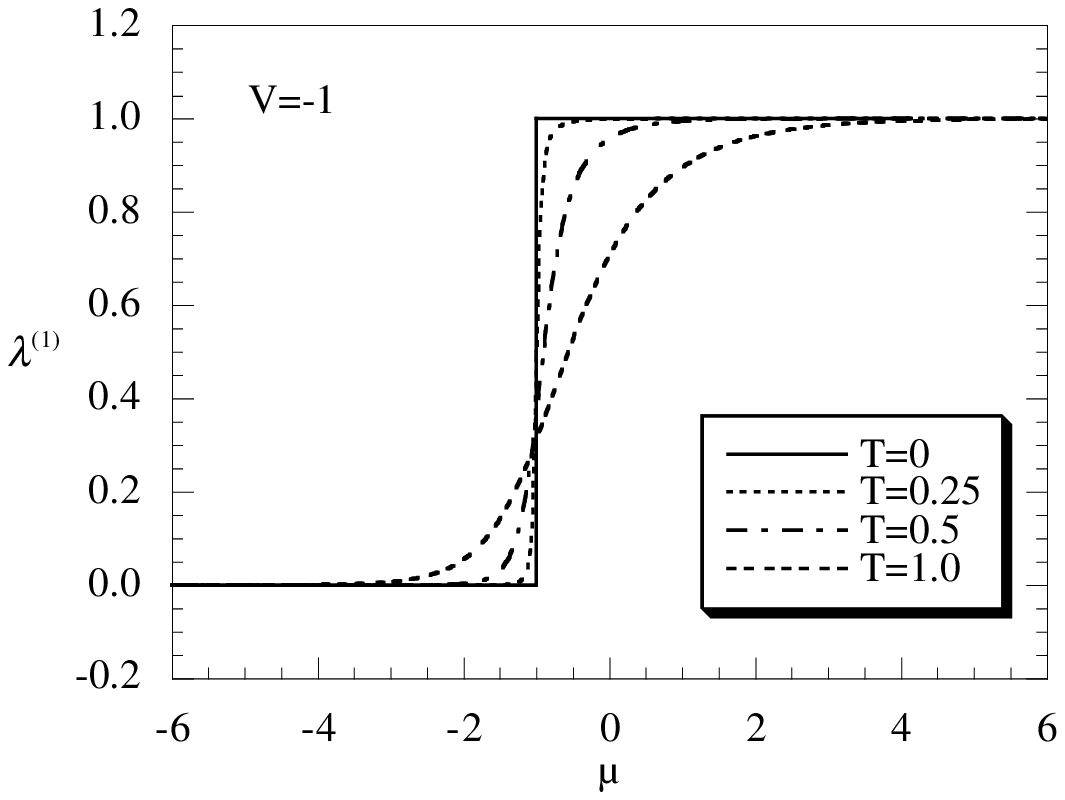} \includegraphics[width=8cm]{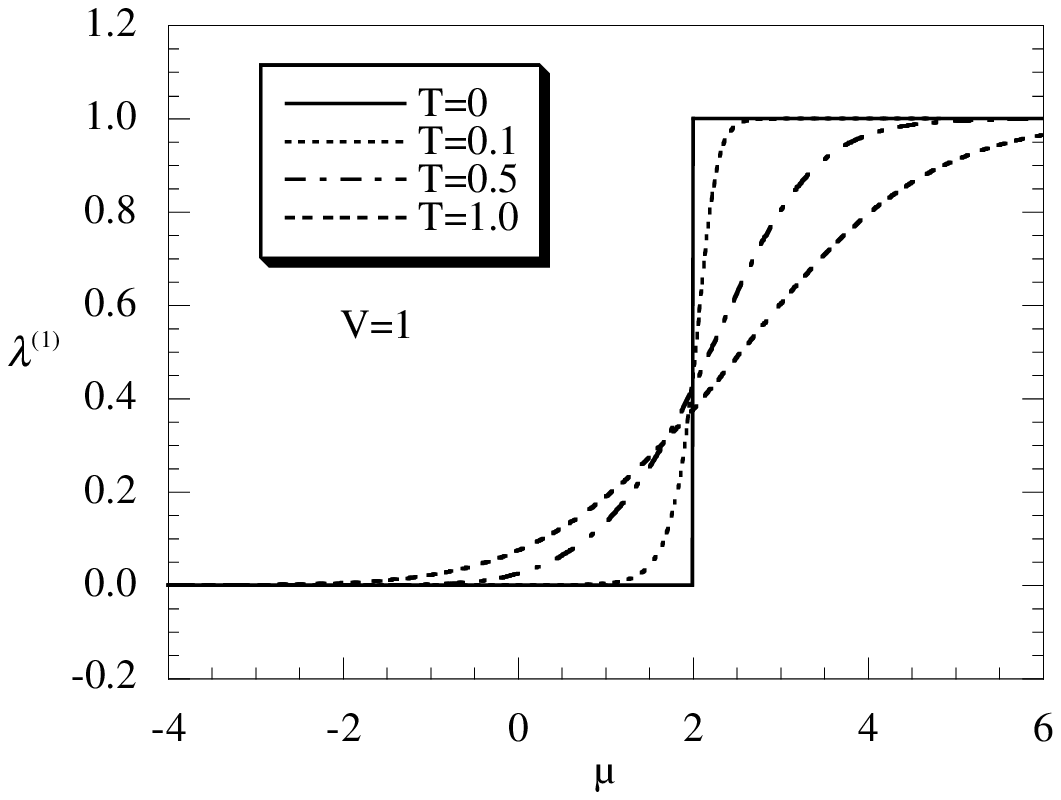}
\caption{The correlation function $\lambda^{(1)}$ is plotted as a
function of the chemical potential at various temperatures for
$V=-1$ (top) and $V=1$ (bottom).} \label{Fig2}
\end{figure}

In Fig.~\ref{Fig2} we give the parameter $\lambda ^{(1)}$ as a
function of $ \mu $.\ For the ferromagnetic case the behavior is
similar to that of $\nu $ .\ Instead, in the antiferromagnetic case
$\lambda ^{(1)}$, at $T=0$, exhibits only one discontinuity point at
$\mu =2V$, where jumps from zero to one. The parameter $\lambda
^{(2)}$ has a behavior similar to $\lambda ^{(1)} $. The different
behavior exhibited by the pairs $(\nu ,\kappa ^{(2)}) $ and
$(\lambda ^{(1)},\lambda ^{(2)})$ for $V<0$ is naturally due to the
antiferromagnetic correlations, when we recall that the pair $
(\lambda ^{(1)},\lambda ^{(2)})$ describes a correlation between two
first neighboring sites, while $\kappa ^{(2)}$ describes
correlations between two second neighboring sites. Of course in the
point of discontinuity the two limits $T\rightarrow 0$ and $\mu
\rightarrow \mu _{c}$ are not interchangeable. As we shall see in
the next Section, the 4 local parameters $\nu ,\kappa ^{(2)},\lambda
^{(1)},\lambda ^{(2)}$ are really basic since all the properties of
the model are described in terms of them. It is worthwhile to note
that some simple relations can be established among the parameters
\begin{equation}
2\kappa ^{(2)}-\nu -\nu ^{2}=\frac{(\nu ^{2}-\lambda ^{(1)})^{2}}{\nu {\
(1-\nu )}}  \label{5.27}
\end{equation}
\begin{equation}
2(\kappa ^{(2)}-\lambda ^{(2)})=(\nu -\lambda ^{(1)})+{\frac{{(\nu -\lambda
^{(1)})^{2}}}{{1-}\nu }}  \label{5.28}
\end{equation}

The Ising model in one dimension can be described in terms of only two
parameters: $\nu $ and $\lambda ^{(1)}$.

\section{Charge correlation functions for one-dimensional systems}

The system of equations (\ref{6.38}) establishes some relations
among the non-local charge correlation functions
${K^{(p)}(\mathbf{i,j})=<[v^{\alpha }(
\mathbf{i})]^{p}v(\mathbf{j})>}$ and ${\Lambda
^{(p)}(\mathbf{i,j})=<v( \mathbf{i})[v^{\alpha
}(\mathbf{i})]^{p}v(\mathbf{j})>}$. As already discussed, the number
of equations is not sufficient to determine completely the charge
correlation functions, and one needs more equations to close the
system. In the one-dimensional case a fourth equation can be easily
obtained by algebraic considerations. Recalling that $c^{\dag
}(i)e^{-\beta H}=c^{\dag }(i)e^{-\beta H_{0}}$ we can easily derive
the following result
\begin{equation}
\Lambda ^{(p)}(i,j)=K^{(p)}(i,j)-C_{1,1}\left\langle [v^{\alpha
}(i)]^{p}v(j)\right\rangle _{0}  \label{7.1}
\end{equation}
Now, for $j>i+a$ (because of invariance under axis reflection we could
choose $j<i-a$ as well), using the property (\ref{5.17})
\begin{eqnarray}
\!\left\langle \nu ^{\alpha }(i)\nu (j)\right\rangle _{0} &=&{\frac{1}{2}}
\left\langle \nu ^{\alpha }(i)\right\rangle _{0}  \nonumber \\
&&\times \left\langle \nu (j)\right\rangle _{0}+{\frac{1}{2}}\left\langle
\nu (i+a)\nu (j)\right\rangle _{0}  \label{7.2}
\end{eqnarray}
\begin{equation}
\left\langle \nu (i+a)\nu (i-a)\nu (j)\right\rangle _{0}=\left\langle \nu
^{\alpha }(i)\right\rangle _{0}\left\langle \nu (i+a)\nu (j)\right\rangle
_{0}  \label{7.3}
\end{equation}
Therefore, from (\ref{7.1})
\begin{equation}
\Lambda ^{(0)}(i,j)=K^{(0)}(i,j)-C_{1,1}\left\langle \nu (j)\right\rangle
_{0}  \label{7.4}
\end{equation}
\begin{eqnarray}
\Lambda ^{(1)}(i,j) &=&K^{(1)}(i,j)-{\frac{1}{2}}C_{1,2}\left\langle \nu
(j)\right\rangle _{0}  \nonumber \\
&&-{\frac{1}{2}}C_{1,1}\left\langle \nu (i+a)\nu (j)\right\rangle _{0}
\label{7.5}
\end{eqnarray}
\begin{eqnarray}
\Lambda ^{(2)}(i,j) &=&K^{(2)}(i,j)-{\frac{1}{2}}[K^{(1)}(i,j)-\Lambda
^{(1)}(i,j)]  \nonumber \\
&&-{\frac{1}{2}}C_{1,2}\left\langle \nu (i+a)\nu (j)\right\rangle _{0}
\label{7.6}
\end{eqnarray}
where we used Eq. (\ref{5.15}). Equations (\ref{7.4}) and (\ref{7.5}) give
\begin{equation}
\left\langle \nu (j)\right\rangle _{0}={\frac{1}{{C_{1,1}}}}
[K^{(0)}(i,j)-\Lambda ^{(0)}(i,j)]  \label{7.7}
\end{equation}
\begin{align}
\left\langle \nu (i+a)\nu (j)\right\rangle _{0} &={\frac{2}{{C_{1,1}}}}
[K^{(1)}(i,j)-\Lambda ^{(1)}(i,j)]  \nonumber \\
&-{\frac{{C_{12}}}{{C_{1,1}^{2}}}}[K^{(0)}(i,j)-\Lambda ^{(0)}(i,j)]
\label{7.8}
\end{align}
Putting (\ref{7.8}) into (\ref{7.6}) we get the fourth self-consistent
equation
\begin{eqnarray}
\Lambda ^{(2)}(i,j) &=&K^{(2)}(i,j)  \nonumber \\
&&-[{\frac{1}{2}}+{\frac{{C_{1,2}}}{{C_{1,1}}}}][K^{(1)}(i,j)-\Lambda
^{(1)}(i,j)]  \nonumber \\
&&+{\frac{{C_{1,2}^{2}}}{{2C_{1,1}^{2}}}}[K^{(0)}(i,j)-\Lambda ^{(0)}(i,j)]
\label{7.9}
\end{eqnarray}
It is straightforward to verify that (\ref{7.9}) is identically satisfied
for $j=i$, while for $j=i^{\alpha }$ coincides with equation (\ref{5.28}).

By adding (\ref{7.9}) to (\ref{6.38}), we have the following system
of equations for the non-local spin correlation functions
$K^{(p)}(i,j)$ and $ \Lambda ^{(p)}(i,j)$
\begin{equation}
{2\Lambda ^{(0)}(i,j)-a_{0}K^{(1)}(i,j)+a_{1}K^{(2)}(i,j)=a_{5}}\nu
\label{7.10}
\end{equation}
\begin{equation}
{2\Lambda ^{(1)}(i,j)+a_{2}K^{(1)}(i,j)-2a_{3}K^{(2)}(i,j)=0}  \label{7.11}
\end{equation}
\begin{equation}
{2\Lambda ^{(2)}(i,j)+a_{3}K^{(1)}(i,j)-a_{4}K^{(2)}(i,j)=0}  \label{7.12}
\end{equation}
\begin{equation}
\begin{array}{l}
{\Lambda ^{(2)}(i,j)+p_{0}[K^{(1)}(i,j)-\Lambda ^{(1)}(i,j)]} \\
{+p_{1}\Lambda ^{(0)}(i,j)-K^{(2)}(i,j)=p_{1}}\nu
\end{array}
\label{7.13}
\end{equation}
where we put
\begin{equation}
\begin{array}{l}
{{a_{0}=(3T_{1}-4T_{2}+T_{3})}} \\
{{a_{1}=2(T_{1}-2T_{2}+T_{3})}} \\
{{a_{2}=(2T_{2}-T_{3}-2)}}
\end{array}
\qquad
\begin{array}{l}
{{a_{3}=(T_{2}-T_{3})}} \\
{{a_{4}=(2+T_{2}-2T_{3})}\hfill } \\
{{a_{5}=(2-T_{1})}}
\end{array}
\label{7.14}
\end{equation}
\begin{equation}
\begin{array}{l}
{{p_{0}=({\frac{1}{2}}+{\frac{{C_{1,2}}}{{C_{1,1}}}})}} \\
{{p_{1}={\frac{{C_{1,2}^{2}}}{{2C_{1,1}^{2}}}}}}
\end{array}
\label{7.15}
\end{equation}
and we used $K^{(0)}(i,j)=\left\langle \nu (j)\right\rangle =\nu $. To
calculate for general $\left| {i-j}\right| $ we proceed by induction.

Let us put $i-j=ma$ and concentrate the attention on the spin CF $\Lambda
^{(0)}(m)=\left\langle \nu (m)\nu (0)\right\rangle $. We start by observing
that
\begin{equation}
{\Lambda ^{(0)}(0)=\nu \qquad \Lambda ^{(0)}(1)=\lambda ^{(1)}}  \label{7.16}
\end{equation}
where the two parameters $\nu ,$ $\lambda ^{(1)}$ have been calculated in
the fermionic sector. By taking $m=2$ we can calculate from the system (\ref
{7.10})-(\ref{7.13}) that
\begin{equation}
\Lambda ^{(0)}(2)=2\kappa ^{(2)}-\nu  \label{7.17}
\end{equation}
The results (\ref{7.16})-(\ref{7.17}) and use of the relation (\ref{5.27})
show that $\Lambda ^{(0)}(m)$ for $m=0,1,2$ can be cast in the form
\begin{equation}
\Lambda ^{(0)}(m)=\nu ^{2}+\nu (1-\nu )p^{m}  \label{7.18}
\end{equation}
where the parameter $p$ is defined as
\begin{equation}
p={\frac{{2\kappa ^{(2)}-\nu -\nu ^{2}}}{{\lambda ^{(1)}-\nu
^{2}}}}={\frac{{ \ \lambda ^{(1)}-\nu ^{2}}}{\nu {-\nu ^{2}}}}
\label{7.19}
\end{equation}
By using the expressions of {the basic parameters given in Section 5, it is
possible to check that }$\left| p\right| <1$.\ Then, we can introduce the
Fourier transform and reexpress (\ref{7.18}) as
\begin{equation}
\Lambda ^{(0)}(m)=\nu ^{2}+A\nu (1-\nu ){\frac{a}{{2\pi }}}\int\limits_{-\pi
/a}^{\pi /a}{}dk{\frac{{e^{ikma}}}{{1+B\cos (ka)}}}  \label{7.20}
\end{equation}
where
\begin{equation}
A={\frac{\nu {-\kappa ^{(2)}}}{{\kappa ^{(2)}-\nu ^{2}}}}\quad \quad B=-{\
\frac{{\ \lambda ^{(1)}-\nu ^{2}}}{{\kappa ^{(2)}-\nu ^{2}}}}  \label{7.21}
\end{equation}
Then, the CF $\Lambda ^{(0)}(m+1)$ can be calculated as
\begin{eqnarray}
{\Lambda ^{(0)}(m+1)} &=&\nu {^{2}+A\nu (1-\nu ){\frac{a}{{2\pi }}}
\int\limits_{-\pi /a}^{\pi /a}{}dk{\frac{{e^{ika(m+1)}}}{{1+B\cos (ka)}}}}
\nonumber \\
{} &=&\nu {^{2}+\nu (1-\nu )p^{m+1}}  \label{7.22}
\end{eqnarray}
Therefore, Eq. (\ref{7.18}) is valid for any $m$.\ Recalling the definition
of $\Lambda ^{(0)}(m)$, we can rewrite (\ref{7.18}) under the form
\begin{equation}
{\frac{{<\nu (m)\nu (0>-\nu ^{2}}}{\nu {-\nu ^{2}}}}=p^{m}  \label{7.23}
\end{equation}
Also, from (\ref{7.20}) we see that the zero frequency function [cfr. (\ref
{3.0a})] has the expression
\begin{equation}
\Gamma (k)=\nu ^{2}(2\pi /a)\delta (k)+{\frac{A\nu (1-\nu )}{{1+B\cos (ka)}}}
\label{7.23a}
\end{equation}

By putting the obtained expression of $\Lambda ^{(0)}(m)$ in Eqs. (\ref{7.10}
)-(\ref{7.13}), we can solve the system. The solution gives
\begin{equation}
K^{(1)}(m)=\nu ^{2}+{\frac{1}{2}}\nu (1-\nu )(p^{m-1}+p^{m+1})  \label{7.24}
\end{equation}
\begin{eqnarray}
K^{(2)}(m) &=&\kappa ^{(2)}\nu +\nu (1-\nu )  \nonumber \\
&&\left[ {{\frac{{a_{0}}}{{2a_{1}}}}(p^{m-1}+p^{m+1})-{\frac{2}{{a_{1}}}}
p^{m}}\right]  \label{7.25}
\end{eqnarray}
\begin{align}
\Lambda ^{(1)}(m) &=\lambda ^{(1)}\nu +\nu (1-\nu ) \times  \nonumber \\
&\times \left[
{{\frac{{2a_{0}a_{3}-a_{1}a_{2}}}{{4a_{1}}}}(p^{m-1}+p^{m+1})- {\
\frac{{2a_{3}}}{{a_{1}}}}p^{m}}\right]  \label{7.26}
\end{align}
\begin{align}
\Lambda ^{(2)}(m) &=\lambda ^{(2)}\nu +\nu (1-\nu ) \times  \nonumber \\
&\times \left[
{{\frac{{a_{0}a_{4}-a_{1}a_{3}}}{{4a_{1}}}}(p^{m-1}+p^{m+1})-{
\frac{{a_{4}}}{{a_{1}}}}p^{m}}\right]  \label{7.27}
\end{align}

\begin{figure}[ptb]
\includegraphics[width=8cm]{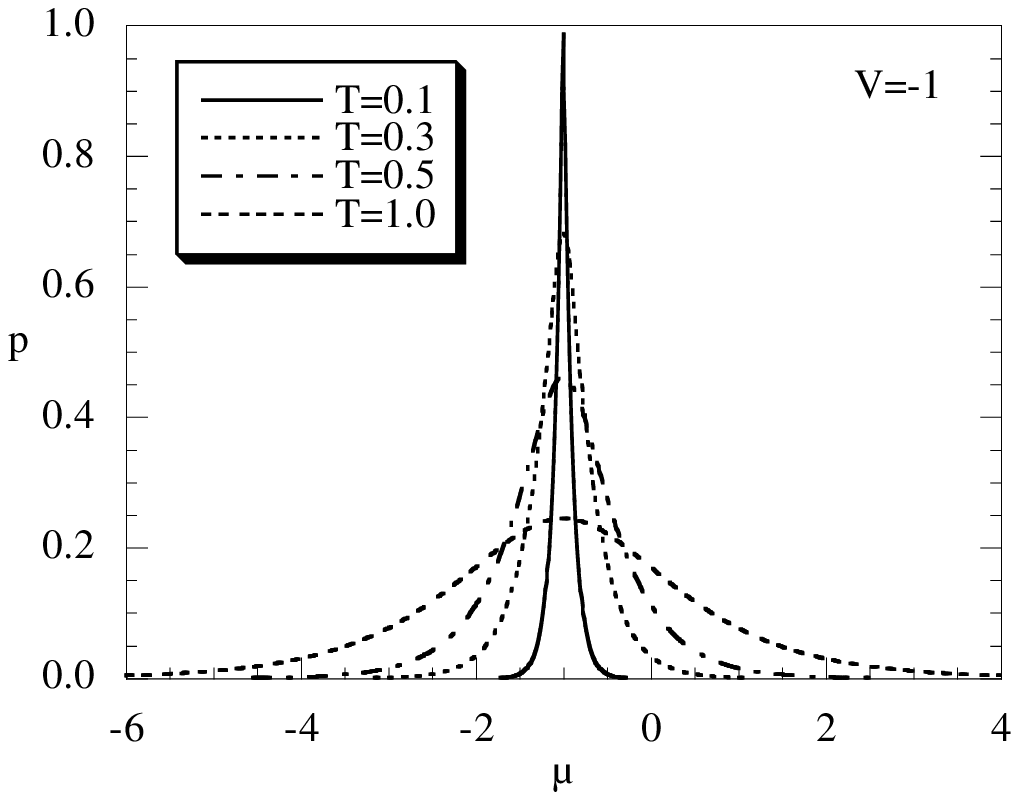} \includegraphics[width=8cm]{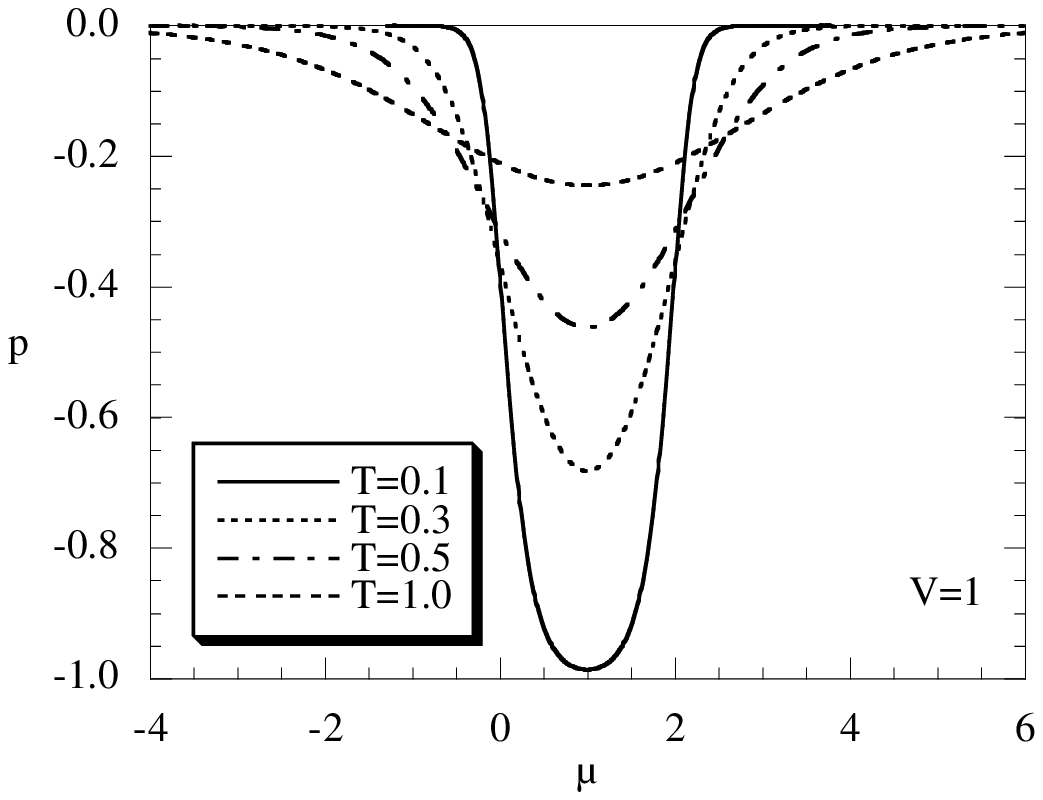}
\caption{The parameter $p$ is plotted as a function of the chemical
potential at various temperatures for $V=-1$ (top) and $V=1$
(bottom).} \label{Fig3}
\end{figure}

\begin{figure}[ptb]
\includegraphics[width=8cm]{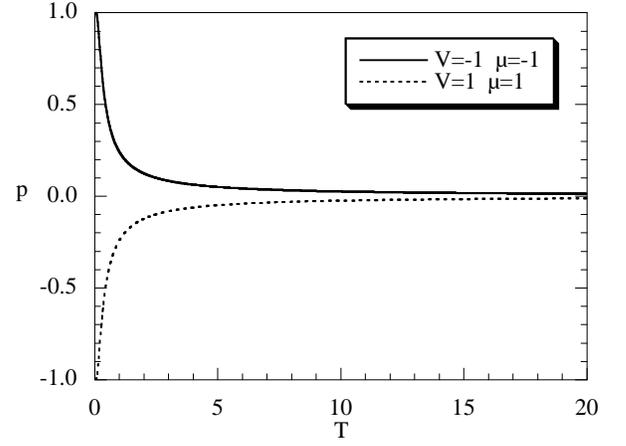}
\caption{The parameter $p$ is plotted as a function of the
temperature at $ \mu=V=-1$ and $\mu=V=1$.} \label{Fig4}
\end{figure}

In Fig.~\ref{Fig3} we give $p$ as a function of $\mu $ for $V=-1$
and $V=1$ at various temperatures. We see that for negative $V$
(ferromagnetic case), $ p$ is positive and various between zero and
1. For positive $V$ (antiferromagnetic case), $p$ is negative and
various between $-1$ and zero. In particular, for negative $V$, $p$
tends to 1 at $\mu =V$ in the limit $ T\to 0$. Instead, for positive
$V$, $p$ tends to $-1$ at $\mu =V$ in the limit $T\to 0$. This is
seen in Fig.~\ref{Fig4} where $p$ is plotted versus $ T$ at $\mu
=V=-1$ and at $\mu =V=1$.

\begin{figure}[ptb]
\includegraphics[width=8cm]{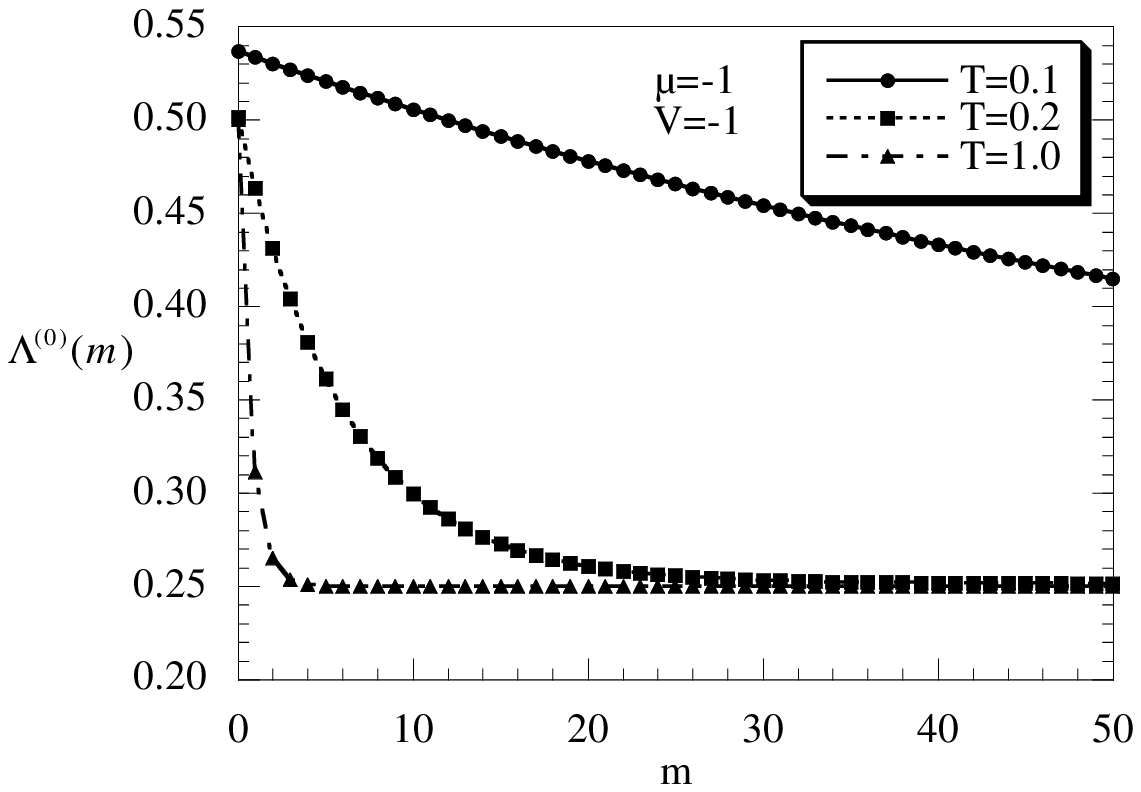} \includegraphics[width=8cm]{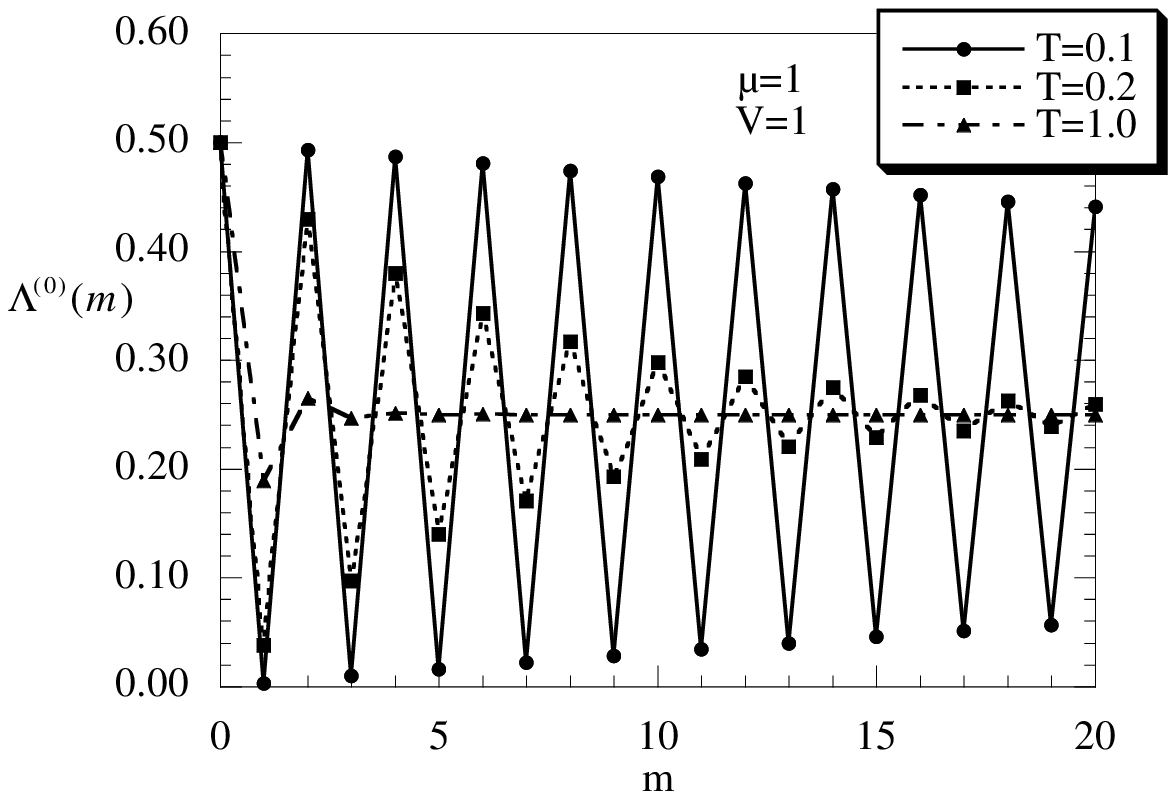}
\caption{$\Lambda^{(0)}(m)$ is plotted against $m$ for $\mu=V=-1$
(top) and $ \mu=V=1$ (bottom) and various temperature.} \label{Fig5}
\end{figure}

Let us now discuss the correlation functions. $\Lambda ^{(0)}(m)$ is
plotted against $m$ for $\mu =V=-1$ [Fig.~\ref{Fig5} (top)] and for
$\mu =V=1$ [Fig.~\ref{Fig5} (bottom)] at various temperatures. We
see that when zero temperature is approached a long-range order of
ferromagnetic and antiferromagnetic type is established,
respectively. Also, we can see from ( \ref{7.22})-(\ref{7.27}) that
for $m\rightarrow \infty $ and $T\neq 0$ (i.e. not at the critical
temperature) the spin correlation functions assume the ergodic
value. At the critical temperature $T=0$ we have breakdown of the
ergodicity.

\section{Conclusions}

The Ising model in presence of an external magnetic field is
isomorphic to a model of localized spinless interacting particles,
satisfying the Fermi statistics. The latter model belongs to a class
of models always solvable, as shown in Ref.~\onlinecite{Mancini05}.
On this basis, we have constructed a general solution of the Ising
model which holds for any dimensionality of the system. The
Hamiltonian of the model has been solved in terms of a complete
finite set of eigenoperators and eigenvalues. The Green's function
and the correlation functions of the fermionic model are exactly
known and are expressed in terms of a finite small number of
parameters that have to be self-consistently determined. By using
the equation of the motion method, we have derived a set of
equations which connect different spin correlation functions. The
scheme that emerges is that it is possible to describe the Ising
model from a unified point of view where all the properties are
connected to a small number of local parameters, and where the
critical behavior is controlled by the energy scales fixed by the
eigenvalues of the Hamiltonian. The latter considerations have been
proved from a quantitative point of view in the one-dimensional
case, where the equations which determine the self-consistent
parameters and the spin correlation function have been solved. For
$d=1$ all the properties of the system have been calculated and
obviously agree with the exact results reported in the literature.
Extension of the calculations to higher dimensions is under
investigation.

After the paper was completed and submitted for publication, the
author learned that approaches to the spin-1/2 Ising model based on
a fermionization of the model have been previously reported in
Refs.~\onlinecite {Tyablikov67} and \onlinecite{Kalashnikov69a}. The
author wishes to thank the referee and Prof.~L.~De Cesare for
putting these papers to his attention. In particular, in
Ref.~\onlinecite{Tyablikov67}, Tyablikov and Fedyanin showed that
the chain of equations for the double-time GF closes and the number
of equations is determined only by the co-ordination number,
independently by the dimensionality. This conclusion agrees with the
results given in Sections 3 and 4. In order to close the set of
equations for the fermionic correlation functions in the
one-dimensional case, the authors of Ref.~\onlinecite{Tyablikov67}
assumed ergodicity and solved the system, obtaining the exact
solution of the 1D Ising model for an infinite chain. It should be
remarked that ergodicity breaks down for finite systems and at the
critical points. Kalashnikov and Fradkin in Ref.~\onlinecite
{Kalashnikov69a} used the spectral density method
\cite{Kalashnikov69} to derive a system of equations for the
correlation functions; also in this case the approach is valid for
any dimension. However, the number of equations is less than the
number of correlation functions.

\appendix

\section{Algebraic relations}

As mentioned in Section 3, the number density operator $\nu (i)=c^{\dag
}(i)c(i)$ satisfies the algebra
\begin{equation}
\nu ^{p}(i)=\nu (i) \quad p\ge 1  \label{A1}
\end{equation}
 From this algebra an important relation can be derived. for the operator
\begin{equation}
\nu ^{\alpha }(i)=\sum\limits_{\mathbf{j}}{}\alpha _{\mathbf{ij}}\nu
(j)={ \frac{1}{{2d}}}\sum\limits_{m=1}^{2d}{}\nu (i_{m})  \label{A2}
\end{equation}
where $i_{m}$ are the first neighbors of the site $\mathbf{i}$. We shall
discuss separately the cases of different dimensions.

\subsection{One dimension}

We start from the equation
\begin{equation}
\lbrack \nu ^{\alpha }(i)]^{p}={\frac{1}{{2^{p}}}}\sum\limits_{m=0}^{p}
\left(
\begin{array}{c}
p \\
m
\end{array}
\right) \nu (i_{1})^{p-m}\nu (i_{2})^{m}  \label{A3}
\end{equation}
After subtracting the terms $m=0$ and $m=p$, we can use the algebraic
relation (\ref{A1}) to obtain
\begin{equation}
\lbrack \nu ^{\alpha }(i)]^{p}={\frac{1}{{2^{p}}}}[2\nu ^{\alpha
}(i)+a_{p}\nu (i_{1})\nu (i_{2})]  \label{A4}
\end{equation}
with
\begin{equation}
a_{p}=\sum\limits_{m=1}^{p-1}\left(
\begin{array}{c}
p \\
m
\end{array}
\right) =2^{p}-2  \label{A5}
\end{equation}
 From (\ref{A4}), by putting $p=2$ we obtain
\begin{equation}
\nu (i_{1})\nu (i_{2})=2[\nu ^{\alpha }(i)]^{2}-\nu ^{\alpha }(i)  \label{A6}
\end{equation}
By substituting (\ref{A6}) into (\ref{A4}) we have the recurrence rule
\begin{equation}
\lbrack \nu ^{\alpha }(i)]^{p}=\sum\limits_{m=1}^{2}{}A_{m}^{(p)}[\nu
^{\alpha }(i)]^{m}  \label{A7}
\end{equation}
where
\begin{eqnarray}
{A_{1}^{(p)}} &=&{{\frac{1}{{2^{p}}}}(2-a_{p})=2^{2-p}-1}  \label{A8} \\
{A_{2}^{(p)}} &=&{{\frac{1}{{2^{p}}}}2a_{p}=2(1-2^{1-p})}
\end{eqnarray}
We note that the coefficients $A_{m}^{(p)}$ satisfy the relation
\begin{equation}
\sum\limits_{m=1}^{2}{}A_{m}^{(p)}=1  \label{A9}
\end{equation}
In table 1 we give the values of the $A_{m}^{(p)}$'s for $1\le p\le 6$.

\begin{tabular}{|c|c|c|}
\hline
p & $A_{1}^{(p)}$ & $A_{2}^{(p)}$ \\ \hline
1 & 1 & 0 \\ \hline
2 & 0 & 1 \\ \hline
3 & $-\frac{1}{2}$ & $\frac{3}{2}$ \\ \hline
4 & $-\frac{3}{4}$ & $\frac{7}{4}$ \\ \hline
5 & $-\frac{7}{8}$ & $\frac{15}{8}$ \\ \hline
6 & $-\frac{15}{16}$ & $\frac{31}{16}$ \\ \hline
\end{tabular}

\subsection{Two dimensions}

We start from the equation
\begin{align}
\lbrack \nu ^{\alpha }(i)]^{p} &={\frac{1}{{4^{p}}}}\sum\limits_{m=0}^{p}
\left(
\begin{array}{c}
p \\
m
\end{array}
\right) \nu (i_{1})^{p-m}\sum\limits_{n=0}^{m}\left(
\begin{array}{c}
m \\
n
\end{array}
\right)  \nonumber \\
&\times \nu (i_{2})^{m-n}\sum\limits_{l=0}^{n}\left(
\begin{array}{c}
n \\
l
\end{array}
\right) \nu (i_{3})^{n-l}\nu (i_{4})^{l}  \label{A10}
\end{align}
By proceeding as in the case of one dimension, use of the algebraic relation
(\ref{A1}) leads to
\begin{equation}
\lbrack \nu ^{\alpha }(i)]^{p}={\frac{1}{{4^{p}}}}\sum
\limits_{m=1}^{4}{}b_{m}^{(p)}Z_{m}  \label{A11}
\end{equation}
where the operators $Z_{m}$ are defined as
\begin{equation}
\begin{array}{l}
{Z_{1}=4\nu ^{\alpha }(i)} \\
{Z_{2}=\nu (i_{1})\nu (i_{2})+\nu (i_{1})\nu (i_{3})+\nu (i_{1})\nu (i_{4})}
\\
{+\nu (i_{2})\nu (i_{3})+\nu (i_{2})\nu (i_{4})+\nu (i_{3})\nu (i_{4})} \\
{Z_{3}=\nu (i_{1})\nu (i_{2})\nu (i_{3})+\nu (i_{1})\nu (i_{2})\nu (i_{4})}
\\
{+\nu (i_{1})\nu (i_{3})\nu (i_{4})+\nu (i_{2})\nu (i_{3})\nu (i_{4})} \\
{Z_{4}=\nu (i_{1})\nu (i_{2})\nu (i_{3})\nu (i_{4})}
\end{array}
\label{A12}
\end{equation}
and the coefficients $b_{m}^{(p)}$ have the expressions
\begin{equation}
\begin{array}{l}
{b_{1}^{(p)}=4} \\
{b_{2}^{(p)}=\sum\limits_{m=1}^{p-1}}\left(
\begin{array}{l}
p \\
m
\end{array}
\right) {=2^{p}-2} \\
{b_{3}^{(p)}=\sum\limits_{m=2}^{p-1}}\left(
\begin{array}{l}
p \\
m
\end{array}
\right) \sum\limits_{n=1}^{m-1}\left(
\begin{array}{l}
m \\
n
\end{array}
\right) {=3(1-2^{p}+3^{p-1})} \\
{b_{4}^{(p)}=\sum\limits_{m=3}^{p-1}}\left(
\begin{array}{l}
p \\
m
\end{array}
\right) \sum\limits_{n=2}^{m-1}\left(
\begin{array}{l}
m \\
n
\end{array}
\right) {\sum\limits_{l=1}^{n-1}{}}\left(
\begin{array}{l}
n \\
n
\end{array}
\right) \\
{=4(-1+3\cdot 2^{p-1}-3^{p}+4^{p-1})}
\end{array}
\label{A13}
\end{equation}
By solving the system (\ref{A11}) with respect to variables $Z_{m}$, we can
obtain the recurrence rule
\begin{equation}
\lbrack \nu ^{\alpha }(i)]^{p}=\sum\limits_{m=1}^{4}{}A_{m}^{(p)}[\nu
^{\alpha }(i)]^{m}  \label{A14}
\end{equation}
where the coefficients $A_{m}^{(p)}$are defined as
\begin{equation}
\begin{array}{l}
{A_{1}^{(p)}=4^{1-p}-2^{1-2p}b_{2}^{(p)}+{\frac{1}{3}}
4^{1-p}b_{3}^{(p)}-4^{-p}b_{4}^{(p)}} \\
{A_{2}^{(p)}=2^{3-2p}b_{2}^{(p)}-2^{3-2p}b_{3}^{(p)}+{\frac{{11}}{3}}
2^{1-2p}b_{4}^{(p)}} \\
{A_{3}^{(p)}={\frac{1}{3}}2^{5-2p}b_{3}^{(p)}-4^{2-p}b_{4}^{(p)}} \\
{A_{4}^{(p)}={\frac{1}{3}}2^{5-2p}b_{4}^{(p)}}
\end{array}
\label{A15}
\end{equation}
We note that for all p
\begin{equation}
\sum\limits_{m=1}^{4}{}A_{m}^{(p)}=1  \label{A16}
\end{equation}
In table 2 we give the values of the $A_{m}^{(p)}$'s for $1\le p\le 8$

\begin{tabular}{|c|c|c|c|c|}
\hline
$p$ & $A_{1}^{(p)}$ & $A_{2}^{(p)}$ & $A_{3}^{(p)}$ & $A_{4}^{(p)}$ \\ \hline
$1$ & $1$ & $0$ & $0$ & $0$ \\ \hline
$2$ & $0$ & $1$ & $0$ & $0$ \\ \hline
$3$ & $0$ & $0$ & $1$ & $0$ \\ \hline
$4$ & $0$ & $0$ & $0$ & $1$ \\ \hline
$5$ & $-\frac{3}{32}$ & $\frac{25}{32}$ & $-\frac{35}{16}$ & $\frac{5}{2}$
\\ \hline
$6$ & $-\frac{15}{64}$ & $\frac{119}{64}$ & $-\frac{75}{16}$ &
$\frac{65}{16} $ \\ \hline $7$ & $-\frac{195}{512}$ &
$\frac{1505}{512}$ & $-\frac{1799}{256}$ & $\frac{ 175}{32}$ \\
\hline $8$ & $-\frac{525}{1024}$ & $\frac{3985}{1024}$ &
$-\frac{1155}{128}$ & $ \frac{1701}{256}$ \\ \hline
\end{tabular}

\subsection{Three dimensions}

We start from the equation
\begin{align}
{\lbrack \nu ^{\alpha }(i)]^{p}} &={{\frac{1}{{6^{p}}}}\sum\limits_{m=0}^{p}
}\left(
\begin{array}{l}
p \\
m
\end{array}
\right) \nu {(i_{1})^{p-m}\sum\limits_{n=0}^{m}}\left(
\begin{array}{l}
m \\
n
\end{array}
\right)  \nonumber \\
&\times \nu {(i_{2})^{m-n}\sum\limits_{l=0}^{n}}\left(
\begin{array}{l}
n \\
l
\end{array}
\right) \nu {(i_{3})^{n-l}}  \label{A17} \\
&\times {\sum\limits_{k=0}^{l}}\left(
\begin{array}{l}
l \\
k
\end{array}
\right) \nu {(i_{4})^{l-k}\sum\limits_{q=0}^{k}}\left(
\begin{array}{l}
k \\
q
\end{array}
\right) \nu {(i_{5})^{k-q}\nu (i_{6})^{q}}  \nonumber
\end{align}
Because of the algebraic relations (\ref{A1}) we obtain
\begin{equation}
\lbrack \nu ^{\alpha }(i)]^{p}={\frac{1}{{4^{p}}}}\sum
\limits_{m=1}^{6}{}b_{m}^{(p)}Z_{m}  \label{A18}
\end{equation}
where the operators $Z_{m}$ are defined as
\begin{align}
{Z_{1}} &={6\nu ^{\alpha }(i)}  \label{A19} \\
{Z_{2}} &={\ \nu _{1}\nu _{2}+\nu _{1}\nu _{3}+\nu _{2}\nu }_{3}{+\nu
_{1}\nu _{4}+\nu _{2}\nu _{4}+\nu _{3}\nu }_{4}  \nonumber \\
&{+\nu _{1}\nu }_{5}{+\nu _{2}\nu _{5}+\nu _{3}\nu }_{5}{+\nu
_{4}\nu }_{5}{
\ +\nu _{1}\nu }_{6}  \nonumber \\
&{+\nu _{2}{\nu }_{6}+\nu _{3}{\nu }_{6}+\nu _{4}{\nu }_{6}+\nu _{5}\nu }
_{6}  \label{A19a} \\
{Z_{3}} &={\ \ \nu _{1}\nu _{2}{\nu }_{3}+\nu _{1}\nu _{2}\nu _{4}+\nu
_{1}\nu _{3}\nu _{4}+\nu _{2}\nu _{3}\nu _{4}+\nu _{1}\nu _{2}\nu _{5}}
\nonumber \\
&{+\nu _{1}\nu _{3}\nu _{5}+\nu _{2}\nu _{3}\nu _{5}+\nu _{1}\nu _{4}\nu
_{5}+\nu _{2}\nu _{4}\nu _{5}+\nu _{3}\nu _{4}\nu _{5}}  \nonumber \\
&{+\nu _{1}\nu _{2}{\nu }_{6}+\nu _{1}\nu _{3}{\nu }_{6}+\nu _{2}\nu _{3}{\
\nu }_{6}+\nu _{1}\nu _{4}{\nu }_{6}+\nu _{2}\nu _{4}\nu }_{6}  \nonumber \\
&{+\nu _{3}\nu _{4}{\nu }_{6}+{\nu }_{1}\nu _{5}{\nu }_{6}+\nu
_{2}\nu _{5}{ \ \nu }_{6}+\nu _{3}\nu _{5}{\nu }_{6}+\nu _{4}\nu
_{5}\nu }_{6}
\label{A19b} \\
{Z_{4}} &={\ \nu _{1}\nu _{2}\nu _{3}\nu _{4}+\nu _{1}\nu _{2}\nu _{3}\nu
_{5}+\nu _{1}\nu _{2}\nu _{4}\nu _{5}+\nu _{1}\nu _{3}\nu _{4}\nu _{5}}
\nonumber \\
&{+\nu _{2}\nu _{3}\nu _{4}\nu _{5}+\nu _{1}\nu _{2}\nu _{3}{\nu }_{6}+\nu
_{1}\nu _{2}\nu _{4}{\nu }_{6}+\nu _{1}\nu _{3}\nu _{4}\nu }_{6}  \nonumber
\\
&{+\nu _{2}\nu _{3}\nu _{4}{\nu }_{6}+\nu _{1}\nu _{2}\nu _{5}{\nu }
_{6}+\nu _{1}\nu _{3}\nu _{5}{\nu }_{6}+\nu _{2}\nu _{3}\nu _{5}\nu }_{6}
\nonumber \\
&{+\nu _{1}\nu _{4}\nu _{5}{\nu }_{6}+\nu _{2}\nu _{4}\nu _{5}{\nu }
_{6}+\nu _{3}\nu _{4}\nu _{5}\nu }_{6}  \label{A19c} \\
{Z_{5}} &={\ \nu _{1}\nu _{2}\nu _{3}\nu _{4}\nu _{5}+\nu _{1}\nu _{2}\nu
_{3}\nu _{4}{\nu }_{6}+\nu _{1}\nu _{2}\nu _{3}\nu _{5}\nu }_{6}  \nonumber
\\
&{+\nu _{1}\nu _{2}\nu _{4}\nu _{5}{\nu }_{6}+\nu _{1}\nu _{3}\nu _{4}\nu
_{5}{\nu }_{6}+\nu _{2}\nu _{3}\nu _{4}\nu _{5}\nu }_{6}  \nonumber \\
{Z_{6}} &={\nu _{1}\nu _{2}\nu _{3}\nu _{4}\nu _{5}\nu }_{6}  \label{A19d}
\end{align}
and the new coefficients $b_{m}^{(p)}(p=5,6)$ have the expressions
\begin{equation}
\begin{array}{l}
{b_{5}^{(p)}=\sum\limits_{m=2}^{p-1}}\left(
\begin{array}{l}
p \\
m
\end{array}
\right) \sum\limits_{n=1}^{m-1}\left(
\begin{array}{l}
m \\
n
\end{array}
\right) {a_{p-m}a_{m-n}} \\
=5(1-2^{p+1}+2\cdot 3^{p}-4^{p}+5^{p-1}) \\
{b_{6}^{(p)}=\sum\limits_{m=2}^{p-1}}\left(
\begin{array}{l}
p \\
m
\end{array}
\right) \sum\limits_{n=1}^{m-1}\left(
\begin{array}{l}
m \\
n
\end{array}
\right) {a_{p-m}a_{m-n}a}_{n} \\
=-6+15\cdot 2^{p}+2^{3+2p}-20\cdot 3^{p}+7\cdot 4^{p}-6\cdot 5^{p}+6^{p}
\end{array}
\label{A21}
\end{equation}
By solving the system (\ref{A18}) with respect to variables $Z_{m}$, we can
obtain the recursion rule
\begin{equation}
\lbrack \nu ^{\alpha }(i)]^{p}=\sum\limits_{m=1}^{6}{}A_{m}^{(p)}[\nu
^{\alpha }(i)]^{m}  \label{A22}
\end{equation}
where the coefficients $A_{m}^{(p)}$ are defined as
\begin{equation}
\begin{array}{l}
{A_{1}^{(p)}={\frac{1}{{6^{p}}}}[6-3b_{2}^{(p)}+2b_{3}^{(p)}-{\frac{3}{2}}
b_{4}^{(p)}+{\frac{6}{5}}b_{5}^{(p)}-b_{6}^{(p)}]} \\
{A_{2}^{(p)}={\frac{1}{{6^{p}}}}[18b_{2}^{(p)}-18b_{3}^{(p)}+{\frac{{33}}{2}}
b_{4}^{(p)}-15b_{5}^{(p)}+{\frac{{137}}{{10}}}b_{6}^{(p)}]} \\
{A_{3}^{(p)}={\frac{1}{{6^{p}}}}[36b_{3}^{(p)}-54b_{4}^{(p)}+63b_{5}^{(p)}-{
\ \frac{{135}}{2}}b_{6}^{(p)}]} \\
{A_{4}^{(p)}={\frac{1}{{6^{p}}}}
[54b_{4}^{(p)}-108b_{5}^{(p)}+153b_{6}^{(p)}] } \\
{A_{5}^{(p)}={\frac{1}{{6^{p}}}}[{\frac{{324}}{5}}
b_{5}^{(p)}-162b_{6}^{(p)}] } \\
{A_{6}^{(p)}={\frac{1}{{6^{p}}}}{\frac{{324}}{5}}b_{6}^{(p)}}
\end{array}
\label{A23}
\end{equation}
We note that for all p
\begin{equation}
\sum\limits_{m=1}^{4}{}A_{m}^{(p)}=1  \label{A24}
\end{equation}
In table 3 we give the values of the $A_{m}^{(p)}$'s for $1\le p\le 10$.

\begin{tabular}{|c|c|c|c|c|c|c|}
\hline $p$ & $A_{1}^{(p)}$ & $A_{2}^{(p)}$ & $A_{3}^{(p)}$ &
$A_{4}^{(p)}$ & $ A_{5}^{(p)}$ & $A_{6}^{(p)}$ \\ \hline $1$ & $1$ &
$0$ & $0$ & $0$ & $0$ & $0$ \\ \hline $2$ & $0$ & $1$ & $0$ & $0$ &
$0$ & $0$ \\ \hline $3$ & $0$ & $0$ & $1$ & $0$ & $0$ & $0$ \\
\hline $4$ & $0$ & $0$ & $0$ & $1$ & $0$ & $0$ \\ \hline $5$ & $0$ &
$0$ & $0$ & $0$ & $1$ & $0$ \\ \hline $6$ & $0$ & $0$ & $0$ & $0$ &
$0$ & $1$ \\ \hline $7$ & $-\frac{5}{324}$ & $\frac{49}{216}$ &
$-\frac{203}{162}$ & $\frac{245}{ 172}$ & $-\frac{175}{36}$ &
$\frac{7}{2}$ \\ \hline $8$ & $-\frac{35}{648}$ &
$\frac{1009}{1296}$ & $-\frac{2695}{648}$ & $\frac{ 13811}{1296}$ &
$-\frac{245}{18}$ & $\frac{133}{18}$ \\ \hline $9$ &
$-\frac{665}{5832}$ & $\frac{6307}{3888}$ & $-\frac{98915}{11664}$ &
$ \frac{9065}{432}$ & $-\frac{10913}{432}$ & $\frac{49}{4}$ \\
\hline $10$ & $-\frac{245}{1296}$ & $\frac{62167}{23328}$ &
$-\frac{53375}{3888}$ & $\frac{774575}{23328}$ & $-\frac{4165}{108}$
& $\frac{7609}{432}$ \\ \hline
\end{tabular}

\section{The energy matrix}

The energy matrix $\epsilon ^{(d)}$, defined by Eq. (\ref{3.7}) can
be immediately calculated by means of the equation of motion
(\ref{3.2}) and the recurrence rule (\ref{3.4}) [see Tables 1, 2,
3]. The matrix $\Omega ^{(d)}$ is defined as the matrix whose
columns are the eigenvectors of the matrix $\epsilon ^{(d)}$. In
this Appendix we report the expressions of $ \epsilon ^{(d)}$ and
$\Omega ^{(d)}$ for the various dimensions.

\subsection{One dimension}

\begin{equation}
\epsilon ^{(1)}=\left(
\begin{array}{ccc}
-\mu & 2V & 0 \\
0 & -\mu & 2V \\
0 & -V & 3V-\mu
\end{array}
\right) \;\;\;\;\;\;\;\Omega ^{(1)}=\left(
\begin{array}{ccc}
1 & 2^{2} & 1 \\
0 & 2 & 1 \\
0 & 1 & 1
\end{array}
\right)  \label{B1}
\end{equation}

\subsection{Two dimensions}

\begin{equation}
\epsilon ^{(2)}=\left(
\begin{array}{ccccc}
-\mu & 4V & 0 & 0 & 0 \\
0 & -\mu & 4V & 0 & 0 \\
0 & 0 & -\mu & 4V & 0 \\
0 & 0 & 0 & -\mu & 4V \\
0 & -\frac{3}{8}V & \frac{25}{8}V & -\frac{35}{4}V & 10V-\mu
\end{array}
\right)  \label{B2}
\end{equation}
\begin{equation}
\Omega ^{(2)}=\left(
\begin{array}{ccccc}
1 & 4^{4} & 2^{4} & (\frac{4}{3})^{4} & 1 \\
0 & 4^{3} & 2^{3} & (\frac{4}{3})^{3} & 1 \\
0 & 4^{2} & 2^{2} & (\frac{4}{3})^{2} & 1 \\
0 & 4 & 2 & (\frac{4}{3}) & 1 \\
0 & 1 & 1 & 1 & 1
\end{array}
\right)  \label{B3}
\end{equation}

\subsection{Three dimensions}

\begin{equation}
\epsilon ^{(3)}=\left(
\begin{array}{ccccccc}
-\mu & 6V & 0 & 0 & 0 & 0 & 0 \\
0 & -\mu & 6V & 0 & 0 & 0 & 0 \\
0 & 0 & -\mu & 6V & 0 & 0 & 0 \\
0 & 0 & 0 & -\mu & 6V & 0 & 0 \\
0 & 0 & 0 & 0 & -\mu & 6V & 0 \\
0 & 0 & 0 & 0 & 0 & -\mu & 6V \\
0 & -\frac{5}{54}V & \frac{49}{36}V & -\frac{203}{27}V & \frac{245}{12}V & -
\frac{175}{6}V & 21V-\mu
\end{array}
\right)  \label{B4}
\end{equation}
\begin{equation}
\Omega ^{(3)}=\left(
\begin{array}{ccccccc}
1 & 6^{6} & 3^{6} & 2^{6} & \left( \frac{3}{2}\right) ^{6} & \left( \frac{6}{
5}\right) ^{6} & 1 \\
0 & 6^{5} & 3^{5} & 2^{5} & \left( \frac{3}{2}\right) ^{5} & \left( \frac{6}{
5}\right) ^{5} & 1 \\
0 & 6^{4} & 3^{4} & 2^{4} & \left( \frac{3}{2}\right) ^{4} & \left( \frac{6}{
5}\right) ^{4} & 1 \\
0 & 6^{3} & 3^{3} & 2^{3} & \left( \frac{3}{2}\right) ^{3} & \left( \frac{6}{
5}\right) ^{3} & 1 \\
0 & 6^{2} & 3^{2} & 2^{2} & \left( \frac{3}{2}\right) ^{2} & \left( \frac{6}{
5}\right) ^{2} & 1 \\
0 & 6 & 3 & 2 & \left( \frac{3}{2}\right) & \left( \frac{6}{5}\right) & 1 \\
0 & 1 & 1 & 1 & 1 & 1 & 1
\end{array}
\right)  \label{B5}
\end{equation}

\section{The normalization matrix}

We recall the definition of the normalization matrix
\begin{eqnarray}
I^{(d)}(\mathbf{i,j}) &=&\left\langle \{\psi ^{(d)}(\mathbf{i},t),\psi
^{(d)}{}^{\dag }(\mathbf{j},t)\}\right\rangle  \nonumber \\
&=&{\frac{1}{N}}\sum\limits_{\mathbf{k}}{}e^{\text{i}\mathbf{k}\cdot
( \mathbf{R}_{i}- \mathbf{R}_{j})}I^{(d)}(\mathbf{k})  \label{C1}
\end{eqnarray}
It is straightforward to see that use of the hermiticity condition (\ref
{4.14} ) leads to the fact that we have to calculate only the matrix
elements $I_{1,m}^{(d)}(\mathbf{k})\;(m=1,2,\cdots 2d+1)$. The calculations
of these is very easy when one observes the following anticommutating rule
\begin{equation}
\begin{array}{l}
\{c(\mathbf{i},t)[\nu ^{\alpha }(i)]^{p},c^{\dag
}(\mathbf{j},t)\}=\delta _{
\mathbf{ij }}[\nu ^{\alpha }(i)]^{p} \\
-\sum\limits_{n=1}^{p}{}(-1)^{n}{\frac{1}{{(2d)^{n-1}}}}\left(
\begin{array}{c}
p \\
n
\end{array}
\right) \alpha _{\mathbf{ij}}c(\mathbf{i},t)[\nu ^{\alpha }(i)]^{p-n}c^{\dag
}( \mathbf{j},t)
\end{array}
\label{C2}
\end{equation}
By taking the expectation value of (\ref{C2}) we obtain in momentum space
\begin{eqnarray}
I_{1,m}^{(d)}(\mathbf{k}) &=&\kappa ^{(m-1)}-\alpha (\mathbf{k})\sum
\limits_{n=1}^{m-1}{}(-1)^{n}  \nonumber \\
&&\times {\frac{1}{{(2d)^{n-1}}}}\left(
\begin{array}{c}
m-1 \\
n
\end{array}
\right) C_{1,m-n}^{(d)\alpha }  \label{C3}
\end{eqnarray}

with the definitions
\begin{equation}
{C^{(d)\alpha }=\left\langle {\psi ^{(d)\alpha }(i)\psi ^{(d)\dag }(i)}
\right\rangle \;\;\qquad \kappa ^{(p)}=\left\langle {[\nu ^{\alpha }(i)]^{p}}
\right\rangle }  \label{C4}
\end{equation}

\subsection{One dimension}

\begin{equation}
I^{(1)}(\mathbf{k})=\left(
\begin{array}{ccc}
I_{1,1}^{(1)} & I_{1,2}^{(1)} & I_{1,3}^{(1)} \\
I_{1,2}^{(1)} & I_{1,3}^{(1)} & I_{2,3}^{(1)} \\
I_{1,3}^{(1)} & I_{2,3}^{(1)} & I_{3,3}^{(1)}
\end{array}
\right)  \label{C5}
\end{equation}

where
\begin{equation}
\begin{array}{l}
I_{1,1}^{(1)}{(\mathbf{k})=1} \\
I_{1,m}^{(1)}(\mathbf{k})=\kappa ^{(m-1)}-\alpha (\mathbf{k})\sum
\limits_{n=1}^{m-1}{}(-1)^{n}{\frac{1}{{(2)^{n-1}}}} \\
\times \left(
\begin{array}{c}
m-1 \\
n
\end{array}
\right) C_{1,m-n}^{(d)\alpha }\qquad (m=2,3) \\
I_{m,3}^{(1)}(\mathbf{k})=\sum
\limits_{n=1}^{2}{}A_{n}^{(m+1)}I_{1,n+1}^{(1)}( \mathbf{k})\qquad
(m=2,3)
\end{array}
\label{C6}
\end{equation}

\subsection{Two dimensions}

\begin{equation}
I^{(2)}(\mathbf{k})=\left(
\begin{array}{ccccc}
I_{1,1}^{(2)} & I_{1,2}^{(2)} & I_{1,3}^{(2)} & I_{1,4}^{(2)} & I_{1,5}^{(2)}
\\
I_{1,2}^{(2)} & I_{1,3}^{(2)} & I_{1,4}^{(2)} & I_{1,5}^{(2)} & I_{2,5}^{(2)}
\\
I_{1,3}^{(2)} & I_{1,4}^{(2)} & I_{1,5}^{(2)} & I_{2,5}^{(2)} & I_{3,5}^{(2)}
\\
I_{1,4}^{(2)} & I_{1,5}^{(2)} & I_{2,5}^{(2)} & I_{3,5}^{(2)} & I_{4,5}^{(2)}
\\
I_{1,5}^{(2)} & I_{2,5}^{(2)} & I_{3,5}^{(2)} & I_{4,5}^{(2)} & I_{5,5}^{(2)}
\end{array}
\right)  \label{C7}
\end{equation}

where
\begin{equation}
\begin{array}{l}
I_{1,1}^{(2)}{{(\mathbf{k})=1}} \\
I_{1,m}^{(2)}(\mathbf{k})=\kappa ^{(m-1)}-\alpha (\mathbf{k})\sum
\limits_{n=1}^{m-1}{}(-1)^{n}{\frac{1}{{(4)^{n-1}}}} \\
\times \left(
\begin{array}{c}
m-1 \\
n
\end{array}
\right) C_{1,m-n}^{(d)\alpha }\qquad (m=2,\cdots 5) \\
I_{m,5}^{(2)}(\mathbf{k})=\sum
\limits_{n=1}^{4}{}A_{n}^{(m+3)}I_{1,n+1}^{(2)}( \mathbf{k})\qquad
(m=2,\cdots 5)
\end{array}
\label{C8}
\end{equation}

\subsection{Three dimensions}

\begin{equation}
I^{(3)}(\mathbf{k})=\left(
\begin{array}{ccccccc}
I_{1,1}^{(3)} & I_{1,2}^{(3)} & I_{1,3}^{(3)} & I_{1,4}^{(3)} & I_{1,5}^{(3)}
& I_{1,6}^{(3)} & I_{1,7}^{(3)} \\
I_{1,2}^{(3)} & I_{1,3}^{(3)} & I_{1,4}^{(3)} & I_{1,5}^{(3)} & I_{1,6}^{(3)}
& I_{1,7}^{(3)} & I_{2,7}^{(3)} \\
I_{1,3}^{(3)} & I_{1,4}^{(3)} & I_{1,5}^{(3)} & I_{1,6}^{(3)} & I_{1,7}^{(3)}
& I_{2,7}^{(3)} & I_{3,7}^{(3)} \\
I_{1,4}^{(3)} & I_{1,5}^{(3)} & I_{1,6}^{(3)} & I_{1,7}^{(3)} & I_{2,7}^{(3)}
& I_{3,7}^{(3)} & I_{4,7}^{(3)} \\
I_{1,5}^{(3)} & I_{1,6}^{(3)} & I_{1,7}^{(3)} & I_{2,7}^{(3)} & I_{3,7}^{(3)}
& I_{4,7}^{(3)} & I_{5,7}^{(3)} \\
I_{1,6}^{(3)} & I_{1,7}^{(3)} & I_{2,7}^{(3)} & I_{3,7}^{(3)} & I_{4,7}^{(3)}
& I_{5,7}^{(3)} & I_{6,7}^{(3)} \\
I_{1,7}^{(3)} & I_{2,7}^{(3)} & I_{3,7}^{(3)} & I_{4,7}^{(3)} & I_{5,7}^{(3)}
& I_{6,7}^{(3)} & I_{7,7}^{(3)}
\end{array}
\right)  \label{C9}
\end{equation}

where
\begin{equation}
\begin{array}{l}
I_{1,1}^{(3)}{{(\mathbf{k})=1}} \\
I_{1,m}^{(3)}(\mathbf{k})=\kappa ^{(m-1)}-\alpha (\mathbf{k})\sum
\limits_{n=1}^{m-1}{}(-1)^{n}{\frac{1}{{(6)^{n-1}}}} \\
\times \left(
\begin{array}{c}
m-1 \\
n
\end{array}
\right) C_{1,m-n}^{(d)\alpha }\qquad (m=2,\cdots 7) \\
I_{m,7}^{(3)}(\mathbf{k})=\sum
\limits_{n=1}^{6}{}A_{n}^{(m+5)}I_{1,n+1}^{(3)}( \mathbf{k})\qquad
(m=2,\cdots 7)
\end{array}
\label{C10}
\end{equation}

\section{The spectral matrices}

The spectral density matrices $\sigma _{ab}^{(d,n)}(\mathbf{k})$ can be
immediately calculated by means of the knowledge of the matrices $\Omega
^{(d)}$ and $I^{(d)}$ through Eq. (\ref{4.6}).

\subsection{One dimension}

\begin{align}
\sigma ^{(1)} &=\Sigma _{1}\left(
\begin{array}{ccc}
1 & 0 & 0 \\
0 & 0 & 0 \\
0 & 0 & 0
\end{array}
\right) \quad \quad \sigma ^{(2)}=\Sigma _{2}\left(
\begin{array}{ccc}
1 & 2^{-1} & 2^{-2} \\
2^{-1} & 2^{-2} & 2^{-3} \\
2^{-2} & 2^{-3} & 2^{-4}
\end{array}
\right) \quad \quad  \nonumber \\
\sigma ^{(3)} &=\Sigma _{3}\left(
\begin{array}{ccc}
1 & 1 & 1 \\
1 & 1 & 1 \\
1 & 1 & 1
\end{array}
\right)  \label{D1}
\end{align}
where
\begin{equation}
\begin{array}{l}
{\Sigma _{1}=I_{1,1}-3I_{1,2}+2I_{1,3}} \\
{\Sigma _{2}=4(I_{1,2}-I_{1,3})} \\
{\Sigma _{3}=-I_{1,2}+2I_{1,3}}
\end{array}
\label{D2}
\end{equation}

\subsection{Two dimensions}

\begin{eqnarray}
\sigma ^{(1)} &=&\Sigma _{1}\left(
\begin{array}{ccccc}
1 & 0 & 0 & 0 & 0 \\
0 & 0 & 0 & 0 & 0 \\
0 & 0 & 0 & 0 & 0 \\
0 & 0 & 0 & 0 & 0 \\
0 & 0 & 0 & 0 & 0
\end{array}
\right) \quad  \label{D4} \\
\sigma ^{(2)} &=&\Sigma _{2}\left(
\begin{array}{ccccc}
1 & 4^{-1} & 4^{-2} & 4^{-3} & 4^{-4} \\
4^{-1} & 4^{-2} & 4^{-3} & 4^{-4} & 4^{-5} \\
4^{-2} & 4^{-3} & 4^{-4} & 4^{-5} & 4^{-6} \\
4^{-3} & 4^{-4} & 4^{-5} & 4^{-6} & 4^{-7} \\
4^{-4} & 4^{-5} & 4^{-6} & 4^{-7} & 4^{-8}
\end{array}
\right)
\end{eqnarray}
\begin{equation}
\sigma ^{(3)}=\Sigma _{3}\left(
\begin{array}{ccccc}
1 & 2^{-1} & 2^{-2} & 2^{-3} & 2^{-4} \\
2^{-1} & 2^{-2} & 2^{-3} & 2^{-4} & 2^{-5} \\
2^{-2} & 2^{-3} & 2^{-4} & 2^{-5} & 2^{-6} \\
2^{-3} & 2^{-4} & 2^{-5} & 2^{-6} & 2^{-7} \\
2^{-4} & 2^{-5} & 2^{-6} & 2^{-7} & 2^{-8}
\end{array}
\right)  \label{D5}
\end{equation}
\begin{eqnarray}
\sigma ^{(4)} &=&\Sigma _{4}\left(
\begin{array}{ccccc}
1 & \left( \frac{4}{3}\right) ^{-1} & \left( \frac{4}{3}\right) ^{-2} &
\left( \frac{4}{3}\right) ^{-3} & \left( \frac{4}{3}\right) ^{-4} \\
\left( \frac{4}{3}\right) ^{-1} & \left( \frac{4}{3}\right) ^{-2} & \left(
\frac{4}{3}\right) ^{-3} & \left( \frac{4}{3}\right) ^{-4} & \left( \frac{4}{
3}\right) ^{-5} \\
\left( \frac{4}{3}\right) ^{-2} & \left( \frac{4}{3}\right) ^{-3} & \left(
\frac{4}{3}\right) ^{-4} & \left( \frac{4}{3}\right) ^{-5} & \left( \frac{4}{
3}\right) ^{-6} \\
\left( \frac{4}{3}\right) ^{-3} & \left( \frac{4}{3}\right) ^{-4} & \left(
\frac{4}{3}\right) ^{-5} & \left( \frac{4}{3}\right) ^{-6} & \left( \frac{4}{
3}\right) ^{-7} \\
\left( \frac{4}{3}\right) ^{-4} & \left( \frac{4}{3}\right) ^{-5} & \left(
\frac{4}{3}\right) ^{-6} & \left( \frac{4}{3}\right) ^{-7} & \left( \frac{4}{
3}\right) ^{-8}
\end{array}
\right) \quad  \label{D6} \\
\sigma ^{(5)} &=&\Sigma _{5}\left(
\begin{array}{ccccc}
1 & 1 & 1 & 1 & 1 \\
1 & 1 & 1 & 1 & 1 \\
1 & 1 & 1 & 1 & 1 \\
1 & 1 & 1 & 1 & 1 \\
1 & 1 & 1 & 1 & 1
\end{array}
\right)
\end{eqnarray}
with
\begin{equation}
\begin{array}{l}
{\Sigma _{1}=I_{1,1}-{\frac{1}{3}}(25I_{1,2}-70I_{1,3}+80I_{1,4}-32I_{1,5})}
\\
{\Sigma _{2}={\frac{{16}}{3}}(3I_{1,2}-13I_{1,3}+18I_{1,4}-8I_{1,5})} \\
{\Sigma _{3}=-4(3I_{1,2}-19I_{1,3}+32I_{1,4}-16I_{1,5})} \\
{\Sigma _{4}={\frac{{16}}{3}}(I_{1,2}-7I_{1,3}+14I_{1,4}-8I_{1,5})} \\
{\Sigma _{5}=-{\frac{1}{3}}(3I_{1,2}-22I_{1,3}+48I_{1,4}-32I_{1,5})}
\end{array}
\label{D7}
\end{equation}

\subsection{Three dimensions}

\begin{eqnarray}
\sigma ^{(1)} &=&\Sigma _{1}\left(
\begin{array}{cccc}
1 & 0 & \cdots & 0 \\
0 & 0 & \cdots & 0 \\
\vdots & \vdots & \vdots & \vdots \\
0 & 0 & \cdots & 0
\end{array}
\right) \qquad  \label{D9} \\
\sigma ^{(2)} &=&\Sigma _{2}\left(
\begin{array}{cccc}
1 & 6^{-1} & \cdots & 6^{-6} \\
6^{-1} & 6^{-2} & \cdots & 6^{-7} \\
\vdots & \vdots & \vdots & \vdots \\
6^{-6} & 6^{-7} & \cdots & 6^{-12}
\end{array}
\right)
\end{eqnarray}
\begin{eqnarray}
\sigma ^{(3)} &=&\Sigma _{3}\left(
\begin{array}{cccc}
1 & 3^{-1} & \cdots & 3^{-6} \\
3^{-1} & 3^{-2} & \cdots & 3^{-7} \\
\vdots & \vdots & \vdots & \vdots \\
3^{-6} & 3^{-7} & \cdots & 3^{-12}
\end{array}
\right) \qquad  \label{D10} \\
\sigma ^{(4)} &=&\Sigma _{4}\left(
\begin{array}{cccc}
1 & 2^{-1} & \cdots & 2^{-6} \\
2^{-1} & 2^{-2} & \cdots & 2^{-7} \\
\vdots & \vdots & \vdots & \vdots \\
2^{-6} & 2^{-7} & \cdots & 2^{-12}
\end{array}
\right)
\end{eqnarray}

\begin{equation}
\sigma ^{(5)}=\Sigma _{5}\left(
\begin{array}{cccc}
1 & \left( \frac{3}{2}\right) ^{-1} & \cdots & \left( \frac{3}{2}\right)
^{-6} \\
\left( \frac{3}{2}\right) ^{-1} & \left( \frac{3}{2}\right) ^{-2} & \cdots &
\left( \frac{3}{2}\right) ^{-7} \\
\vdots & \vdots & \vdots & \vdots \\
\left( \frac{3}{2}\right) ^{-6} & \left( \frac{3}{2}\right) ^{-7} & \cdots &
\left( \frac{3}{2}\right) ^{-12}
\end{array}
\right)  \label{D11}
\end{equation}

\begin{eqnarray}
\sigma ^{(6)} &=&\Sigma _{6}\left(
\begin{array}{cccc}
1 & \left( \frac{6}{5}\right) ^{-1} & \cdots & \left( \frac{6}{5}\right)
^{-6} \\
\left( \frac{6}{5}\right) ^{-1} & \left( \frac{6}{5}\right) ^{-2} & \cdots &
\left( \frac{6}{5}\right) ^{-7} \\
\vdots & \vdots & \vdots & \vdots \\
\left( \frac{6}{5}\right) ^{-6} & \left( \frac{6}{5}\right) ^{-7} & \cdots &
\left( \frac{6}{5}\right) ^{-12}
\end{array}
\right) \qquad  \label{D12} \\
\sigma ^{(7)} &=&\Sigma _{7}\left(
\begin{array}{cccc}
1 & 1 & \cdots & 1 \\
1 & 1 & \cdots & 1 \\
\vdots & \vdots & \vdots & \vdots \\
1 & 1 & \cdots & 1
\end{array}
\right)
\end{eqnarray}
here
\begin{equation}
\Sigma _{p}=\sum_{m=1}^7 B_{m}^{(p)}I_{1,m}^{(3)}  \label{D13}
\end{equation}

\smallskip The coefficients $B_{m}^{(p)}$are given in Table 4.

\begin{tabular}{|c|c|c|c|c|c|c|c|}
\hline $p$ & $B_{1}^{(p)}$ & $B_{2}^{(p)}$ & $B_{3}^{(p)}$ &
$B_{4}^{(p)}$ & $ B_{5}^{(p)}$ & $B_{6}^{(p)}$ & $B_{7}^{(p)}$ \\
\hline $1$ & $1$ & $-\frac{147}{10}$ & $\frac{406}{5}$ &
$-\frac{441}{2}$ & $315$ & $-\frac{1134}{5}$ & $\frac{324}{5}$ \\
\hline $2$ & $0$ & $36$ & $-\frac{1566}{5}$ & $1044$ & $-1674$ &
$1296$ & $-\frac{ 1944}{5}$ \\ \hline $3$ & $0$ & $-45$ &
$\frac{1053}{2}$ & $-\frac{4149}{2}$ & $3699$ & $-3078$ & $972$ \\
\hline $4$ & $0$ & $40$ & $-508$ & $2232$ & $-4356$ & $3888$ &
$-1296$ \\ \hline $5$ & $0$ & $-\frac{45}{2}$ & $297$ &
$-\frac{2763}{2}$ & $2889$ & $-2754$ & $972$ \\ \hline $6$ & $0$ &
$\frac{36}{5}$ & $-\frac{486}{5}$ & $468$ & $-1026$ & $\frac{ 5184
}{5}$ & $-\frac{1944}{5}$ \\ \hline $7$ & $0$ & $-1$ &
$\frac{137}{10}$ & $-\frac{135}{2}$ & $153$ & $-162$ & $
\frac{324}{5}$ \\ \hline
\end{tabular}

\section{Relations between the Ising and spinless model}

In this Appendix, we want to recall the main results of the Ising model and
establish the relations between the two models. For an infinite chain the
simplest method is the use of the transfer matrix method.\ The details of
calculations are well known and can be found in many textbooks.\ For
example, we refer the reader to Refs.\cite{Baxter82,Goldenfeld92,Lavis99}.
The magnetization per site is given by
\begin{equation}
\left\langle S(i)\right\rangle ={\frac{{\sinh (\beta h)}}{{\sqrt{\sinh
^{2}(\beta h)+e^{-4\beta J}}}}}  \label{E3}
\end{equation}
The two-point correlation function $\left\langle S(i)S(i+j)\right\rangle $
has the expression
\begin{equation}
{}\left\langle {S(i)S(i+j)}\right\rangle {=}\left\langle S(i)\right\rangle
^{2}+(1-\left\langle S(i)\right\rangle ^{2})p^{j}  \label{E4}
\end{equation}
where
\begin{equation}
p={\frac{{\gamma ^{(2)}}}{{\gamma ^{(1)}}}}  \label{E16}
\end{equation}
${\gamma }^{(1)\text{ }}$and ${\gamma ^{(2)}}$ are the eigenvalues of the
transfer matrix
\begin{equation}
\begin{array}{l}
{\gamma ^{(1)}\quad =e^{\beta J}\left[ {\cosh (\beta h)+\sqrt{\sinh
^{2}(\beta h)+e^{-4\beta J}}}\right] } \\
{\gamma }^{(2)}=e^{\beta J}\left[ {\cosh (\beta h)-\sqrt{\sinh ^{2}(\beta
h)+e^{-4\beta J}}}\right]
\end{array}
\label{E17}
\end{equation}

The three-point correlation function $\ $is given by\cite{Marsh66}
\begin{equation}
\begin{array}{l}
\left\langle S(i)S(i+j)S(i+j+r)\right\rangle =\left\langle S(i)\right\rangle
^{3} \\
+\left\langle S(i)\right\rangle [1-\left\langle S(i)\right\rangle
^{2}](p^{j}+p^{r}-p^{j+r})
\end{array}
\label{E24}
\end{equation}

The relations between the Ising and fermionic models are

\begin{equation}
\begin{array}{l}
{h={\frac{1}{2}}(\mu -Vd)} \\
{J=-{\frac{1}{4}}Vd}
\end{array}
\label{E29}
\end{equation}
\begin{equation}
\left\langle {\nu (i)}\right\rangle {={\frac{1}{2}}[1+\left\langle
{S(i)} \right\rangle ]}  \label{E30}
\end{equation}
\begin{equation}
{\lambda ^{(1)}={\frac{1}{4}}[1+2\left\langle {S(i)}\right\rangle
+\left\langle {S(i)S(i+a)}\right\rangle ]}  \label{E31}
\end{equation}
\begin{equation}
{\kappa ^{(2)}={\frac{1}{8}}[3+4\left\langle {S(i)}\right\rangle
+\left\langle {S(i)S(i+2a)}\right\rangle ]}  \label{E32}
\end{equation}
\begin{align}
{\lambda ^{(2)}}& ={{\frac{1}{{16}}}[3+7\left\langle {S(i)}\right\rangle
+4\left\langle {S(i)S(i+a)}\right\rangle }  \nonumber \\
& +\left\langle {S(i)S(i+2a)}\right\rangle {+\left\langle
{S(i)S(i+a)S(i+2a)} \right\rangle ]}  \label{E33}
\end{align}

By recalling (\ref{5.23}) and by means of (\ref{E30}), the magnetization in
the fermionic model has the expression
\begin{equation}
{\left\langle {S(i)}\right\rangle }=(1-T_{2})\sqrt{\frac{T_{1}}{
T_{1}-2T_{1}T_{2}+2T_{2}^{2}}}  \label{E34}
\end{equation}
By oserving that $T_{1}$ and $T_{2}$ can be expressed as
\begin{equation}
\begin{array}{l}
T_{1}=1-\tanh \left( \frac{\beta \mu }{2}\right) =\frac{2e^{-2\beta
h}}{
e^{-2\beta h}+e^{-4\beta J}} \\
T_{2}=1-\tanh \left( \frac{\beta (\mu -V)}{2}\right) =1-\tanh (\beta
h)= \frac{2}{e^{2\beta h}+1}
\end{array}
\label{E35}
\end{equation}
it is straigtforward to see that (\ref{E34}) is the same as (\ref{E3}). In
the fermionic model, by means of (\ref{7.23}), we have
\begin{equation}
\frac{\left\langle \nu (i)\nu (i+j)\right\rangle -\nu ^{2}}{\nu -\nu
^{2}}= \frac{\left\langle S(i)S(i+j)\right\rangle -{\left\langle
{S(i)} \right\rangle }^{2}}{1-{\left\langle {S(i)}\right\rangle
}^{2}}=p^{j} \label{E36}
\end{equation}
where the parameter $p$ is expressed in terms of $\nu $ and $\lambda
^{(1)}$ by means of (\ref{7.19}). By using (\ref{5.21}) and
(\ref{5.23}), and by recalling (\ref{5.22a}) and (\ref{E35}), it is
easy to see that the expression of $p$, given by (\ref{7.19}) is
exactly equal to the expression (\ref{E16}). Then, the two-point
correlation function of the fermionic model exactly agree with the
expression (\ref{E4}) of the Ising model. The parameters $\kappa
^{(2)}$, $\lambda ^{(1)}$ and ${\lambda ^{(2)}}$ can be calculated
in the fermionic model by putting (\ref{5.23}) into (\ref{5.20})-(
\ref{5.22}), and in the Ising model by means of
(\ref{E31})-(\ref{E33}). After lengthy, but straightforward,
calculations, using the relations (\ref{5.22a}) and (\ref{E35}), it
is possible to show that there is an exact agreement.


\newcommand{\noopsort}[1]{} \newcommand{\printfirst}[2]{#1}
  \newcommand{\singleletter}[1]{#1} \newcommand{\switchargs}[2]{#2#1}

\end{document}